\documentclass[letterpaper,11pt]{article}
\usepackage{caption}
\pdfoutput=1 
\usepackage{xcolor}
\usepackage{enumerate}
\usepackage{framed}
\usepackage{mdframed}
\usepackage{amsmath}
\usepackage{amsfonts}
\usepackage{mathrsfs}
\usepackage{amssymb}
\usepackage[normalem]{ulem}
\usepackage{graphicx, rotating}
\usepackage{epsfig}
\usepackage{latexsym}
\usepackage{graphicx}
\usepackage{color}
\usepackage{amsmath,bm,amssymb}
\usepackage{cite}
\usepackage{slashed}
\usepackage{hyperref}
\usepackage{epstopdf}
\usepackage[font=footnotesize,labelfont=]{caption}
\AppendGraphicsExtensions{.tif}
\hypersetup{colorlinks=true, citecolor=bluscuro, linkcolor=black, urlcolor=bluscuro}
\definecolor{rossos}{cmyk}{0,1,1,0.55}
\definecolor{bluscuro}{rgb}{0.15, 0.2, .85}
\definecolor{bluchiaro}{cmyk}{1,.3,0.,0.1}

\newcommand{\be}{\begin{equation}}
\newcommand{\ee}{\end{equation}}
\newcommand{\bea}{\begin{eqnarray}}
\newcommand{\eea}{\end{eqnarray}}
\newcommand{\beas}{\begin{eqnarray*}}
\newcommand{\eeas}{\end{eqnarray*}}

\def\BH{\text{\tiny BH}}
\def\DM{\text{\tiny DM}}
\newcommand{\llp}{\left [}
\newcommand{\rrp}{\right ]}
\newcommand{\lp}{\left (}
\newcommand{\rp}{\right )}
\def\d{{\rm d}}

\begin{document}
\def\thefootnote{\fnsymbol{footnote}}

\begin{center}
\LARGE{\textbf{Superfluid Dark Matter around Black Holes}} \\[0.5cm]
 
\large{Valerio De Luca\footnote{vdeluca@sas.upenn.edu} and Justin Khoury\footnote{jkhoury@sas.upenn.edu}}
\\[0.5cm]

\small{
\textit{Center for Particle Cosmology, Department of Physics and Astronomy, University of Pennsylvania,\\ Philadelphia, PA 19104}}

\vspace{.2cm}

\end{center}

\vspace{.6cm}

\hrule \vspace{0.2cm}
\centerline{\small{\bf Abstract}}
\vspace{0.1cm}
{\small\noindent 
Superfluid dark matter, consisting of self-interacting light particles  that thermalize and condense to form a superfluid in galaxies, provides a novel theory that matches the success of the standard $\Lambda$CDM model on cosmological scales while simultaneously offering a rich phenomenology on galactic scales. Within galaxies, the dark matter density profile consists of a nearly homogeneous superfluid core surrounded by an isothermal envelope. In this work we compute the density profile of superfluid dark matter around supermassive black holes at the center of galaxies. We show that, depending on the fluid equation of state, the dark matter profile presents distinct power-law behaviors, which can be used to distinguish it from the standard results for collisionless dark matter.
}

\vspace{0.3cm}
\noindent
\hrule
\def\thefootnote{\arabic{footnote}}
\setcounter{footnote}{0}

\section{Introduction} 
The standard $\Lambda$ Cold Dark Matter ($\Lambda$CDM) model, in which dark matter (DM) consists of non-relativistic, purely gravitationally-interacting particles, fits extremely well the background expansion history, the shape of the cosmic microwave background and matter power spectra, as well as the abundance and mass function of galaxy clusters, with strong evidence coming from observations on the largest scales. 

However, as simulations and observations of galaxies have improved, a number of challenges appeared when one focuses on galactic scale phenomena~\cite{Bullock:2017xww}. In particular, in the absence of baryons, the numerically predicted Navarro-Frank-White (NFW) density profile~\cite{Navarro:1995iw} in inner regions of galaxies and clusters seems to mismatch the one inferred from observations, giving rise to the so-called ``core-cusp problem". Once baryons are included in simulations~\cite{Oman:2015xda}, they play a crucial role in modifying the cuspy cold dark matter distribution to cored profiles, even though  assumptions have to be made about the stellar evolution history of galaxies. 

Furthermore, striking correlations between the gravitational acceleration of baryons and their distribution in disk galaxies are found~\cite{McGaugh:2016leg,Lelli:2016cui}, which represent a generalized version of the baryonic Tully-Fisher relation connecting the fourth power of the asymptotic circular velocity to the total baryonic mass of disk galaxies~\cite{McGaugh:2000sr,McGaugh:2011ac,Famaey:2011kh,Papastergis:2016jqv}. Much progress has been made to derive such correlations in the context of CDM, though their small scatter remains puzzling~\cite{Lelli:2015wst,DiCintio:2015eeq,Desmond:2016azy}. Alternatively, the observed galactic relations may originate from interactions between baryons and dark matter~\cite{Famaey:2017xou,Famaey:2019baq}.

Straightforward extensions to the standard model consist in endowing DM with self-interactions that are strong enough to modify the galactic core~\cite{Spergel:1999mh,Kaplinghat:2015aga}. Another well-studied extension is to ``fuzz out" the cuspy profiles by considering extremely light DM particles, with mass $m\lesssim 10^{-21}\,{\rm eV}$, called fuzzy dark matter~\cite{Hu:2000ke,Hui:2016ltb}.
These ultra-light particles have a de Broglie wavelength of order kiloparsec, thus behaving as a Bose-Einstein condensate on astrophysical scales. (See~\cite{Ferreira:2020fam} for a review.)

In this paper we instead focus on a third possibility, represented by superfluid dark matter. Over a certain mass range and scattering length, these bosonic particles can support condensation  at high density and low temperature~\cite{Goodman:2000tg}. An enticing phenomenological aspect of superfluid dark matter models is the possibility of phonons mediating long-range interactions between baryons~\cite{Berezhiani:2015pia,Berezhiani:2015bqa,Berezhiani:2017tth,Sharma:2018ydn,Berezhiani:2018oxf}. 
The simplest superfluid theory, with quartic self-interactions, was initially thought to be ruled out in Ref.~\cite{Slepian:2011ev}, using observations of galactic rotation curves and Bullet Cluster constraints~\cite{Markevitch:2003at,Clowe:2003tk}. However, the idea has been revitalized recently in Refs.~\cite{Berezhiani:2021rjs,Sharma:2022jio,Berezhiani:2022buv}, by relaxing the simplifying assumptions of global thermal equilibrium and spherical symmetry made in Ref.~\cite{Slepian:2011ev}.

One of the most compelling features of most galaxies, including the Milky Way~\cite{10.1093/mnras/291.1.219,Eckart:1997em, Ghez:2003qj}, is that they are believed to host a supermassive black hole (SMBH) at their center~\cite{Richstone:1998ky, 10.1007/978-94-011-4750-7_11}. This hypothesis may apply as well to other bound structures containing dark matter, such as galaxy clusters or satellite systems. 
These SMBHs with masses in the range $10^6$--$10^{10}\,M_{\odot}$ are believed to power active galactic nuclei and quasars, and may
be generated from baryonic processes, such as the collapse of Pop III 
stars~\cite{Madau_2001} and supermassive stars~\cite{Shibata_2002}, or primordial black holes~\cite{Bean:2002kx, Serpico:2020ehh, DeLuca:2022bjs},
followed by mergers and baryonic gas accretion, even though their origin is still unknown.

The presence of such black holes is expected to strongly modify the density profile of the DM around it, depending both on the
properties of the DM particles and the formation history of the central black hole. For example, for collisionless DM, assuming that the SMBH grows adiabatically from a smaller seed by gas accretion~\cite{1972GReGr...3...63P}, the BH will perturb the particle orbits, resulting into a DM spike characterized by a power-law behavior with slope $-9/4$ in its outer region~\cite{Gondolo:1999ef}. On the other hand, the gravitational scattering of stars in the inner region can heat up the DM fluid and give rise to a softened profile with slope $-3/2$~\cite{Merritt:2003qk, Gnedin:2003rj, Merritt:2006mt} (see also Ref.~\cite{Shapiro:2022prq} for a recent analysis) or even to disruption~\cite{Wanders:2014xia}, while DM annihilations can result into a smoother cusp, with slope $-1/2$~\cite{Vasiliev:2007vh,Shapiro:2016ypb}.
Other spikes can be obtained for alternative BH formation histories, such as from direct collapse of gas inside DM halos~\cite{Begelman:2006db}, from the growth of inspiralling  seed~\cite{Ullio:2001fb}, or in the presence of DM self-interactions~\cite{Fornasa:2007nr,Shapiro:2014oha,Feng:2021qkj, Chavanis:2019bnu}.
Focusing on the latter, finally, the presence of particle collisions will wash out any initial and/or adiabatically altered particle distribution on a collisional relaxation time scale,
and the cusp will follow the power law behavior with slope $-3/4$ as found in Ref.~\cite{Shapiro:2014oha}.

In this work we will evaluate the profile of superfluid DM around massive black holes. In Sec.~2 we will review the basic formalism of the superfluid DM model. In Section 3 we will calculate the DM density profile in different regions around SMBHs. In Sec.~4 we put these results together to discuss the entire DM density profile and corresponding mass function. Section~5 is devoted to the conclusions. In the following we use natural units $\hbar = c = 1$.

\section{Brief review of superfluid dark matter}

In this section we review the basic notions of the theory of superfluid DM. Following Refs.~\cite{Berezhiani:2015pia,Berezhiani:2015bqa,Berezhiani:2017tth,Sharma:2018ydn,Berezhiani:2021rjs,Sharma:2022jio}, DM forms 
a superfluid inside galaxies with a coherence length of order the size of galaxies. These DM particles should be light and characterized by sufficiently strong self-interactions, such that they thermalize and create a core at the center of galaxies. 

In a theory of self-interacting bosons, a superfluid state may be obtained through Bose-Einstein condensation, which occurs if two requirements are fulfilled. The first requirement amounts to demanding that the de Broglie wavelength,~$\lambda_\text{\tiny dB} \sim 1/mv$, of DM particles overlap in the (cold and dense enough) central region of galaxies. That is,~$\lambda_\text{\tiny dB}$ must be larger than the average interparticle separation,~$d\sim \left(m/\rho\right)^{1/3}$, in terms of the DM density profile $\rho(r)$ of the galaxy and the one-dimensional DM velocity dispersion $v(r)$  before the phase transition. This condition implies an upper bound on the DM mass
\be
m \lesssim \left(\frac{\rho_\text{\tiny V}}{v_\text{\tiny V}^3}\right)^{1/4}\,,
\label{BECcondbasic}
\ee
which is evaluated at virialization. This latter process occurs, according to standard collapse theory, when the overdensity is about~200 times the present critical density, such that~\cite{Peacock:1999ye}
\begin{align}
\nonumber
\rho_\text{\tiny V} &= 200 \frac{3H_0^2}{8\pi G} \simeq 1.95\times 10^{-27} \; {\rm g}/{\rm cm}^3\,; \\
\nonumber
R_\text{\tiny V} &= \left(\frac{3M_\text{\tiny DM}}{4\pi \rho_\text{\tiny V}}\right)^{1/3}  \simeq 203 \left(\frac{M_\text{\tiny DM}}{10^{12}M_\odot}\right)^{1/3}{\rm kpc}\,;\\
v_\text{\tiny V} &= \sqrt{\frac{1}{3} \frac{G M_\text{\tiny DM}}{R_\text{\tiny V}}}\simeq 85\left(\frac{M_\text{\tiny DM}}{10^{12}M_\odot}\right)^{1/3}~{\rm km}/{\rm s}\,,
\label{virialrho}
\end{align}
where we have assumed $H_0 = 70~{\rm km} \, {\rm s}^{-1}{\rm Mpc}^{-1}$ for concreteness, and denoted with $M_\text{\tiny DM}$ the mass of the DM halo. Substituting these expressions into Eq.~\eqref{BECcondbasic} gives
\be
m \lesssim 2.3 \left(\frac{M_\text{\tiny DM}}{10^{12}M_\odot}\right)^{-1/4}\; {\rm eV}\,.
\ee
This condition implies that only light enough particles might have de Broglie wavelenghts which overlap significantly at the center of galaxies, therefore forming a Bose-Einstein condensate, while heavy particles do not. The corresponding  wavefunction would describe the gas of particles as a whole system rather than treating its particles individually. When, at fixed density $\rho$, the system temperature $T$ drops below the critical temperature
\begin{equation}
T_{\rm c} = \frac{2\pi}{m^{5/3}k_{\rm B}}\left(\frac{\rho}{\zeta(3/2)}\right)^{2/3}\,,
\label{T_c ideal}
\end{equation}
where $\zeta$ denotes the Riemann zeta function, and~$k_{\rm B}$ indicates the Boltzmann constant, a macroscopically large number of particles occupy the ground state, giving rise to a Bose-Einstein condensate. We will denote this condition as {\it degeneracy}.

The second condition requires DM {\it thermalization} within galaxies, that is, 
bosons reach their maximal entropy state via interactions. Since high densities facilitate equilibration by enhancing interaction rates, thermalization is expected to be more efficient in the core of the environment rather than in its outskirts. Thermalization is therefore achieved when 
particles within a thermal radius have experienced at least one interaction over the galaxy lifetime $t_g$, which is of order~$13$~Gyr for characteristic galaxies.\footnote{If phase-space reshuffling due to dynamical effects is not negligible, then the dynamical time scale~$t_\text{\tiny dyn} = r/v$ would be more appropriate.}
This gives the condition
\begin{equation}
\Gamma t_g \gtrsim 1\,,
\label{eq:conditionRT}
\end{equation}
where $\Gamma$ denotes the interaction rate. The latter depends on the theory at hand, and can be approximated as~\cite{Sikivie:2009qn}
 \begin{equation}
\Gamma =(1+\mathcal{N})\frac{\sigma}{m}\rho v\,,
 \label{eq:csRate}
\end{equation}
in terms of the scattering cross section~$\sigma$, and the Bose-enhancement factor 
\be
\mathcal{N}=\frac{\rho}{m}\left(\frac{2\pi}{m v}\right)^3\,,
\ee
which accounts for the fact that particles are scattering into a highly degenerate phase space. Equation~\eqref{eq:conditionRT} translates into a lower bound on the scattering cross section:
\be
\frac{\sigma}{m} \gtrsim  \left(\frac{m}{{\rm eV}}\right)^{4} \left(\frac{M_\text{\tiny DM}}{10^{12}M_\odot}\right)^{2/3}\; 0.1~\frac{{\rm cm}^2}{{\rm g}}\,.
\ee
For~$m\lesssim {\rm eV}$, this satisfies constraints~\cite{Miralda-Escude:2000tvu,Gnedin:2000ea,Randall:2008ppe} on the cross section of self-interacting DM~\cite{Spergel:1999mh}. 

An upper bound on DM self-interactions comes from the analysis of the Bullet Cluster event~\cite{Markevitch:2003at, Clowe:2003tk}, where in the collision of two clusters the gas component is observed to be displaced with respect to the DM component. In this case, the mean number of scatterings that a DM particle may undergo while passing through the target cluster must be negligible, giving rise to the well-known limit
\be
\frac{\sigma}{m} \lesssim \frac{{\rm cm}^2}{{\rm g}}\,,
\label{bullet}
\ee
under the assumption of non-degeneracy and 2-body scattering.

The superfluid nature of DM dramatically modifies its macroscopic behavior. In particular, instead of behaving as individual collisionless particles, the condensate behaves like a homogeneous classical field configuration with finite number density, characterized by a low-energy spectrum of excitations around the homogeneous condensate, {\it i.e.}, phonon degrees of freedom. Such excitations can mediate an emergent long-range interaction between baryons in the superfluid phase, even though the form of these interactions is model-dependent~\cite{Berezhiani:2018oxf}. 

In the following we will consider two different Lagrangians for the interacting particles, giving rise to different equations of state $P(\rho)$, that are ultimately responsible for the nature of the condensate.

\subsection{Two-body interacting superfluid}
One of the simplest interacting DM models consists of a non-relativistic massive complex scalar field $\psi$ minimally coupled to gravity and characterized by quartic self-interaction of the form
\begin{equation}
\mathcal{H}_2 = \int \text{d}^3x \left[-\frac{1}{2m} \psi^{\dagger}(x)\nabla^2 \psi(x) + \frac{1}{2}g_2 \,\psi^{\dagger\, 2}(x) \psi^2(x)  \right]\,.
\label{H2}
\end{equation}
The parameter $g_2$ controls the strength of the contact interactions. Upon Bose-Einstein condensation this system can exhibit superfluidity, provided that $g_2>0$. At zero temperature, the superfluid equation of state is given by
\begin{equation}
\label{EOS2}
P_2=\frac{g_2 \rho^2}{2m^2} = \frac{2\pi a}{m^3} \rho^2\,,
\end{equation}
where we have introduced the $2 \to 2$ scattering length
\be
a =\frac{m g_2}{4\pi}\,.
\ee
The scattering length can be constrained by the upper limit on the cross section from the Bullet Cluster event (Eq.~\eqref{bullet}). Using~$\sigma = 4 \pi a^2$, this gives
\be
a \lesssim 10^{-7} \sqrt{\frac{m}{\rm \mu eV}} \; {\rm fm}\,.
\ee
The sound speed of phonon excitations is as usual given by
\be
c_s^2 = \frac{\partial P_2}{\partial \rho} = \frac{4\pi a}{m^3} \rho\,.
\label{cs2}
\ee
The non-relativistic approximation inherent in Eq.~\eqref{H2} is valid as long as~$c_s \ll 1$. Furthermore, both the equation of state and sound speed receive thermal corrections proportional to the temperature of the system, suppressed by~$T/T_{\rm c}$~\cite{Sharma:2018ydn}. We ignore such corrections for simplicity.

\subsection{Three-body interacting superfluid}
Going beyond the theory with quartic self-interactions, one can consider superfluids with predominantly 3-body interactions, which is the relevant case for the original DM superfluidity scenario~\cite{Berezhiani:2015bqa}. This corresponding Hamiltonian for a complex scalar field involves a hexic potential\footnote{Our primary motivation for considering~$P\sim \rho^3$ is the DM superfluidity scenario of Ref.~\cite{Berezhiani:2015bqa}. As emphasized in~\cite{Berezhiani:2015bqa}, such an equation of state likely arises from a strongly coupled system. The toy Hamiltonian of Eq.~\eqref{H3body initial} is only intended to illustrate how such equation of state may arise from a microphysical model, and as such we do not consider the bound on its parameters from the Bullet Cluster constraint.}
\begin{equation}
\mathcal{H}_3 = \int {\rm d}^3x \biggl[ -\frac{1}{2m}\psi^{\dagger}(x)\nabla^2 \psi(x) + \frac{1}{3}g_3\, \psi^{\dagger\, 3}(x) \psi^3(x) \biggr] \, ,
\label{H3body initial}
\end{equation}
in terms of the coupling constant $g_3$ controlling the strength of interactions. The corresponding zero-temperature equation of state is given by
\begin{equation}
\label{EOS3}
P_3=\frac{2g_3\rho^3}{3m^3} =  \frac{\rho^3}{12\Lambda^2m^6}\,,
\end{equation}
where we have introduced the cut-off scale $\Lambda$ as
\be
g_3 = \frac{1}{8\Lambda^2 m^3}  \,.
\label{g3 Lambda}
\ee
Following Ref.~\cite{Berezhiani:2015bqa} we have mind the fiducial values~$m = {\rm eV}$ and~$\Lambda = {\rm meV}$ for these parameters.
In this case the sound speed is given by
\be
c_s^2 = \frac{\partial P_3}{\partial \rho} = \frac{\rho^2}{4\Lambda^2m^6}\,.
\label{cs3}
\ee

\section{Density profile around black holes}
In this section we compute the density profile of superfluid DM around a central massive black hole, with mass $M_\BH$ and Schwarzschild radius
\be 
\label{Schwarzschild}
r_\BH = 2 G M_\BH \simeq 9.6 \, \cdot 10^{-7}\, {\rm kpc} \lp \frac{M_\BH}{10^{10} M_\odot}\rp.
\ee
One expects that the presence of the BH will result in enhanced interactions among DM particles, steepening the corresponding density profile from its nearly homogeneous central core with density~$\rho_0$. This modification will occur within 
the  BH sphere of influence, described by a radius $r_h$. This can be estimated as the radius at which the DM potential is equal to the potential induced by the presence of the BH or, equivalently, where the total enclosed DM mass and the BH mass are comparable:
\be
r_h = \lp \frac{3M_\BH}{4 \pi \rho_0} \rp^{1/3} 
\simeq 5.5 \, {\rm kpc} \lp \frac{M_\BH}{10^{10} M_\odot}\rp^{1/3} \lp \frac{\rho_0}{10^{-24} {\rm g/cm^3}} \rp^{-1/3}\,.
\ee
At distances larger than~$r_h$, the BH has negligible effect on the superfluid DM profile.

In order to simplify the nature of the problem, we assume that the 
distribution of DM particles is spherically-symmetric in space,
isotropic in velocity, and has relaxed to a near-equilibrium state. 
We further assume that the BH mass $M_\BH$ is much smaller than the total integrated
DM halo mass~$M_\DM$, while it dominates the total mass of all particles bound to it in the cusp. 

We describe the density profile across three different regions. The outer region, at distances~$r \gtrsim r_h$ from the BH, identifies the range where the BH is not effective enough in modifying the DM profile, and the standard results for superfluid DM apply. Specifically, this region describes the superfluid core with radius $R_{2,3}$ and its match to the standard NFW envelope. 
In the intermediate region, at radii smaller than $r_h$, the BH starts to modify the DM density profile due to its gravitational well, resulting in deviations from the outer profile. This intermediate region continues until relativistic effects to the DM motion become so important that they modify the properties of the superfluid and its equation of state. In particular, in the three-body interacting superfluid, the condition of degeneracy is found to break down at a certain radius $r_\text{\tiny deg}$, after which an inner region starts, further modifying the DM density profile.

\subsection{Outer region}

For distances larger than the BH sphere of influence $r_h$, the DM density profile should approach the standard results found in the literature, see Refs.~\cite{Berezhiani:2015pia,Berezhiani:2015bqa}. In particular, the DM halo is characterized by a superfluid core,
where the condition of degeneracy and thermalization are satisfied,  with the ambient temperature being subcritical (due to high density). Within this region, the presence of repulsive self-interactions is crucial to make the core stable. At distances larger than a critical radius, on the other hand, thermal corrections induced from the environment start to be relevant and likely result in an isothermal envelope. The dynamics on these scales is quite complex, due to possible fragmentation and tidal disruption events that can occur~\cite{Berezhiani:2021rjs,Berezhiani:2022buv}. 

In order to simplify the picture, we will assume that, beyond the superfluid core, the complex distribution of tidal debris can be approximated as an 
NFW profile~\cite{Navarro:1995iw, Navarro:1996gj}, as predicted by $N$-body simulations~\cite{Kaplinghat:2015aga} ignoring the effect of baryons:
\begin{equation}
\rho(r)=\frac{\rho_\text{\tiny NFW}}{\frac{r}{r_s}\left(1+\frac{r}{r_s}\right)^2}\,.
\label{eq:NFW}
\end{equation}
 The characteristic density~$\rho_\text{\tiny NFW}$ and scale parameter~$r_s$ depend on the considered halo. Its size is obtained in terms of the virial radius, $R_\text{\tiny V}$, at which the mean DM density is about 200 times the critical density.  
 The ratio between these scales defines the concentration parameter, $c=R_\text{\tiny V}/r_s$, which is related to the total mass by the mass-to-concentration relation~\cite{Dutton:2014xda}. Typical values of these parameters for a Milky Way-like galaxy are $\rho_\text{\tiny NFW} =  10^{-25} \text{g}/\text{cm}^{3}$, $c= 6$ and $R_\text{\tiny V} = 200$~kpc.

The DM density profile of the condensate halo can be computed 
assuming hydrostatic equilibrium. Under spherical symmetry,  the pressure and acceleration are related by
\be
\frac{1}{\rho (r)} \frac{\d P(r)}{\d r} = - \frac{\d \Phi_\DM(r)}{\d r} - \frac{\d \Phi_\BH (r)}{\d r} = - \frac{4 \pi G}{r^2} \int^r_0 \d r' r'^2\rho(r') - \frac{G M_\BH}{r^2},
\label{hydrostatic 1}
\ee 
where we have assumed the BH potential to be $\Phi_\BH = - G M_\BH/r$. 
The BH sphere of influence~$r_h$ defined above describes the distance at which the DM and BH contributions equal each other in the gravitational potential. 
At larger radii $r \gtrsim r_h$ one can therefore neglect the BH potential contribution on the right-hand side of Eq.~\eqref{hydrostatic 1}, such that
\be
\frac{1}{\rho(r)}\frac{{\rm d}P(r)}{{\rm d}r} \simeq - \frac{4\pi G}{r^2} \int_0^r {\rm d}r' r'^2\rho(r')\,; \qquad r \gtrsim r_h\,.
\label{hydrostatic 2}
\ee
The corresponding profile therefore depends on the equation of state of superfluid DM ({\it i.e.}, $P_2 \propto \rho^2$ or $P_3 \propto \rho^3$) and, as we will show later, will give rise to a superfluid core of size~$R_{2,3}$, respectively.
 
\vspace{0.25cm}
In what follows, we derive the density profile of the superfluid core under different assumptions for the particle self-interactions.

\vspace{0.2cm}
\noindent
{\bf{a) Two-body interacting superfluid:}} In the case where DM particles interact primarily via two-body interactions, the zero-temperature superfluid equation of state is given by $P_2 \sim \rho^2$, as shown in Eq.~\eqref{EOS2}. Substituting this into Eq.~\eqref{hydrostatic 2}, one obtains
\be
\frac{4 \pi a}{m^3} \frac{\d \rho}{\d r} = - \frac{4\pi G}{r^2} \int_0^r {\rm d}r' r'^2\rho(r')\,.
\label{hydrostatic 3}
\ee
It is convenient to rewrite this equation in terms of dimensionless variables~$\Xi_2$ and~$\xi_2$, defined by
\begin{align}
\Xi_2 \equiv \frac{\rho(r)}{\rho_0} \,; \qquad  \xi_2 \equiv \sqrt{\frac{G m^3}{a}} r \,.
\end{align}
Differentiating Eq.~\eqref{hydrostatic 3} with respect to~$r$, and expressing the result in the new variables, it is straightforward to obtain
the $n = 1$ Lane-Emden equation
\be
\frac{1}{\xi^2_2}\frac{{\rm d}}{{\rm d}\xi_2} \left(\xi^2_2 \frac{{\rm d}\Xi_2}{{\rm d}\xi_2}\right) = - \Xi_2\,.
\ee
The boundary conditions that must be imposed are:~$i)$ $\Xi_2(\xi_{2\,h}) = 1$, to ensure
continuity of the density at~$r_h$; and $ii)$ a continuity condition on the derivative 
\begin{align}
\Xi_2'(\xi_{2\,h}) &= - \frac{2}{\xi_{2\,h}^2} \frac{G^{3/2} M_\BH m^{9/2}}{4 \pi \rho_0 a^{3/2}} \nonumber \\
& \simeq -  \frac{2 \cdot 10^{-2}}{\xi_{2\,h}^2}
\lp \frac{M_\BH}{10^{10} M_\odot}\rp
\lp \frac{m}{{\rm \mu eV}} \rp^{9/2} \lp \frac{a}{10^{-11} \, {\rm fm}} \rp^{-3/2} \lp \frac{\rho_0}{10^{-24}\, {\rm g/cm^3}} \rp^{-1}\,,
\end{align}
where the factor of 2 in the first line is obtained from the equality of the DM and BH potential at~$r_h$. The corresponding solution is found to vanish at $\xi_2 = \xi_{2\,h} + \pi$, and approximately given by
\be
\rho(r)\simeq \rho_0 \frac{{\rm sin}\llp \sqrt{\frac{G m^3}{a}}(r-r_h)\rrp}{\sqrt{\frac{G m^3}{a}}(r-r_h)}\,,
\label{rho core2}
\ee
giving rise to a condensate of size
\be
R_2 = \sqrt{\frac{a}{G m^3}}\; (\xi_2 - \xi_{2\,h}) \simeq 55 \, {\rm kpc} \lp \frac{m}{{\rm \mu eV}} \rp^{-3/2} \lp \frac{a}{10^{-11} \, {\rm fm}} \rp^{1/2}\,.
\label{R2}
\ee
In estimating the size of the condensate we have focused on a DM mass and scattering length compatible with the Bullet cluster constraint. 

\vspace{0.2cm}
\noindent
{\bf{b) Three-body interacting superfluid:}} Adopting instead the superfluid equation of state $P_3 \sim \rho^3$, given in Eq.~\eqref{EOS3}, one gets the equation
\be
\frac{1}{8 \Lambda^2 m^6} \frac{\d \rho^2}{\d r} = - \frac{4\pi G}{r^2} \int_0^r {\rm d}r' r'^2\rho(r').
\ee
Introducing the dimensionless variables $\Xi_3$ and $\xi_3$ as
\be
\Xi_3^{1/2}  \equiv \frac{\rho (r)}{\rho_0} \,;\qquad 
\xi_3  \equiv \sqrt{\frac{32\pi G \Lambda^2 m^6}{\rho_0}}~r \,,
\ee
it is straightforward to obtain in this case the~$n = 1/2$ Lane-Emden equation
\be
\frac{1}{\xi^2_3}\frac{{\rm d}}{{\rm d}\xi_3} \left(\xi^2_3 \frac{{\rm d}\Xi_3}{{\rm d}\xi_3}\right) = - \Xi^{1/2}_3\,.
\ee
The boundary conditions are $\Xi_3(\xi_{3\,h}) = 1$, and
\begin{align}
\Xi'_3(\xi_{3\,h}) &= - \frac{2}{\xi_{3\,h}^2} \frac{32 \sqrt{2 \pi} G^{3/2} M_\BH\Lambda^3 m^9 }{\rho_0^{5/2}} \nonumber \\
& \simeq -  \frac{3 \cdot 10^{-2}}{\xi_{3\,h}^2}
\lp \frac{M_\BH}{10^{10} M_\odot}\rp
\lp \frac{m}{\rm eV} \rp^{9} \lp \frac{\Lambda}{{\rm meV}} \rp^{3} \lp \frac{\rho_0}{10^{-24} {\rm g/cm^3}} \rp^{-5/2}.
\end{align}
The numerical solution is found to vanish at $\xi_3 \simeq \xi_{3 \, h} + 2.75$, and can be analytically fitted as
\be
\rho(r)\simeq \rho_0 \cos^{1/2}\left[\frac{\pi (r-r_h)}{2 R_3}\right]\,,
\label{rho core3}
\ee
in terms of the size of the central soliton
\be
\label{Rhalo}
R_3 = \sqrt{\frac{\rho_0}{32\pi G \Lambda^2 m^6}}\; (\xi_3-\xi_{3 \, h}) \simeq 44 \, {\rm kpc} \lp \frac{m}{\rm eV} \rp^{-3} \lp \frac{\Lambda}{\rm meV} \rp^{-1} \lp \frac{\rho_0}{10^{-24} {\rm g/cm^3}} \rp^{1/2}.
\ee
Compared to Eq.~\eqref{R2}, we see that the size of the condensate for three-body interactions is comparable to the two-body case. (Notice, in particular, the different fiducial mass of DM particles in the two cases: ${\rm eV}$ for the cubic case, compared to $\mu$eV for the quadratic interactions.) Another key difference is that~$R_3$ depends on the central density~$\rho_0$, whereas~$R_2$ does not.

\subsection{Intermediate region}
\label{interm reg}

For radii $r \lesssim r_h$, the presence of the BH modifies the density profile due to its gravitational potential. We can describe a region, dubbed ``intermediate'', where the BH mainly dictates the characteristic velocity of gravitationally bound DM particles as
\be
v(r) = \sqrt{\frac{G M_\BH}{r}}\,,
\label{v interm}
\ee
which implies that DM particles at smaller radii move faster.
Assuming energy equipartition, one can ascribe a gas temperature to DM particles via
\be
k_{\rm B} T (r) = \frac{1}{3}m v^2(r)\,,
\ee
such that the temperature profile in this region is given by
\be
T (r) = \frac{1}{3 k_{\rm B}} \frac{G m M_\BH}{r}\,.
\ee
Within this region, we will assume that the BH influence is, however, not strong enough to modify the equation of state of the DM fluid. In the following we will therefore show how the profile changes assuming different equations of state for the superfluid DM.

Since~$r_h$ by definition marks the distance where the DM and BH contributions equal each other, one can make the simplifying assumption that at smaller radii the BH dominates the gravitational potential. The condition of hydrostatic equilibrium, given by Eq.~\eqref{hydrostatic 1}, thus becomes
\be
\frac{\d P(r)}{\d r} \simeq -\rho (r) \frac{G M_\BH}{r^2}\,; \qquad r \lesssim r_h\,.
\label{hydro 3}
\ee 
However, we will find that the non-relativistic approximation breaks down at sufficiently small radii, when DM velocities approach the speed of light. Anticipating this, one must therefore consider the full general relativity treatment. 
Neglecting the small contribution to the stress-energy tensor of DM particles orbiting the black hole, one can adopt the Schwarzschild metric to describe the spherically-symmetric space-time as
\begin{equation}
{\rm d}s^2 = - \left(1-\frac{2G M_\BH}{r}\right) {\rm d}t^2 + \frac{{\rm d} r^2}{1-\frac{2G M_\BH}{r}} + r^2 {\rm d}\Omega^2\,.
\end{equation}
The relativistic generalization of Eq.~\eqref{hydro 3} is then given by~\cite{Shapiro:2014oha}
\be
\frac{\d P(r)}{\d r} = - \frac{\rho (r)+P(r)}{1-2 G M_\BH/r} \frac{G M_\BH}{r^2}\,.
\label{hydro rel}
\ee
The solution to this equation provides therefore the density profile of the superfluid DM in the intermediate region. 

In what follows we will solve Eq.~\eqref{hydro rel} for the two-body and three-body superfluid equations of state. Such treatment is only valid, however, provided that DM is sufficiently cold to be in a superfluid state. If the DM temperature exceeds the critical temperature~$T_{\rm c}$, given by Eq.~\eqref{T_c ideal}, then the condition of degeneracy for Bose-Einstein condensation breaks down. In other words, the assumption of Bose-Einstein degeneracy ($T < T_{\rm c}$) is valid provided that\footnote{We use the non-relativistic expression for~$T_{\rm c}$ for simplicity. As the gas becomes relativistic, Eq.~\eqref{v deg} will receive corrections. For instance, in the ultra-relativistic regime~$T \gg m$, the critical temperature instead scales as~$T_c \propto \varrho^{1/3}$~\cite{Bernstein:1990kf}, in terms of the charge density $\varrho$. However, this regime does not apply to the superfluid particles, which are relativistic but not ultra-relativistic, since at most their velocities become $v \simeq 1/2$ close to the BH accretion radius~$4 G M_\BH$.}   
\be
v^2 < \frac{6 \pi \rho^{2/3}}{m^{8/3} \zeta^{2/3}(3/2)}\,.
\label{v deg}
\ee
With the identification~$v^2 = G M_\BH/r$, degeneracy breaks down at a radius~$r_\text{\tiny deg}$ given by
\be
r_\text{\tiny deg}  \,\rho^{2/3}(r_\text{\tiny deg}) = \frac{G M_\BH m^{8/3} \zeta^{2/3}(3/2)}{6 \pi}\,.
\label{rdeg def}
\ee
For distances $r \lesssim r_\text{\tiny deg}$, which denote the beginning of the ``inner region", the DM component is no longer degenerate, and its equation of state is instead approximated by the ideal gas law,~$P \propto \rho v^2$. 
As we will see below, it may happen that~$r_\text{\tiny deg}$ is smaller than the BH horizon, in which case the DM remains in the superfluid state all the way to the BH horizon. Let us stress that, on the other hand, thermalization never breaks down as we approach the BH.

\vspace{0.25cm}
\noindent
{\bf{a) Two-body interacting superfluid}:} Implementing the equation of state $P_2 \propto \rho^2$, given by Eq.~\eqref{EOS2}, one finds that Eq.~\eqref{hydro rel} becomes
\be
\frac{\d \rho}{\d r} = 
- \frac{1}{1-2 G M_\BH/r} \frac{G M_\BH}{r^2} \left(\frac{m^3}{4 \pi a} + \frac{\rho}{2}\right)\,. 
\ee
This gives the profile
\be
\rho (r) = \frac{m^3}{2\pi a} \left[\left(\frac{1-2GM_\BH/r_h}{1-2GM_\BH/r}\right)^{1/4}-1\right] + \left(\frac{1-2GM_\BH/r_h}{1-2GM_\BH/r}\right)^{1/4}\rho_0 \,,
\label{rho rel}
\ee
where we have fixed the integration constant by requiring that~$\rho (r_h) = \rho_0$. As expected, the DM density increases with decreasing distance to the BH.

The non-relativistic regime corresponds to large distances, $r,r_h \gg 2 G M_\BH$, wherein Eq.~\eqref{rho rel} simplifies to
\be
\label{rhoint2}
\rho (r) = \frac{G M_\BH m^3}{4 \pi a} \lp \frac{1}{r} - \frac{1}{r_h} \rp + \rho_0\,.
\ee
It is easy to see that this is a solution to Eq.~\eqref{hydro 3}. In particular, at short distances the profile becomes a~$1/r$ power-law:
\be
\label{intsim2}
\rho(r) \simeq \frac{G M_\BH m^3}{4 \pi a r}\,.
\ee
As discussed above, this solution is only valid provided that the DM temperature remains below critical. 
Substituting the relativistic density profile of Eq.~\eqref{rho rel} (with $r_h \gg 2 G M_\BH$ and neglecting the subleading contribution proportional to $\rho_0$ at small distances), Eq.~\eqref{v deg} becomes
\be
v^3 < \frac{3 \sqrt{6\pi}}{\zeta (3/2) m a} \llp \frac{1}{(1-2 v^2)^{1/4}} -1\rrp\,.
\ee
In other words, the gas remains degenerate provided that
\be
\frac{\zeta (3/2) m a}{3 \sqrt{6\pi}} < \frac{1}{v^3} \llp \frac{1}{(1-2 v^2)^{1/4}} -1\rrp\,.
\ee
It is easy to see that this condition is always satisfied for the characteristic values of the DM mass and scattering length we focus on, implying that the degeneracy condition holds all the way to the radius $4 G M_\BH$ at which accretion becomes important. 

Incidentally, at the accretion radius the fluid sound speed (Eq.~\eqref{cs2}) is
\be
c_s^2 (r = 4 G M_\BH) = \frac{4\pi a}{m^3} \rho (4 G M_\BH) \simeq 0.39\,,
\ee
showing that we have already approached the relativistic regime at some larger radius. As discussed earlier, for simplicity we neglect corrections to the non-relativistic superfluid equation of state, $P_2 \propto \rho^2$. In this approximation, the profile~\eqref{rho rel} thus remains valid up a radius of order the BH horizon, where it will drop off due to accretion effects. See next subsection for details. 

\vspace{0.2cm}
\noindent
{\bf{b) Three-body interacting superfluid:}} Applying the same procedure to the equation of state in Eq.~\eqref{EOS3}, one finds that Eq.~\eqref{hydro rel} in this case becomes
\be
\frac{\d \rho}{\d r} = 
- \frac{1}{1-2 G M_\BH/r} \frac{G M_\BH }{r^2} 
\lp \frac{4 \Lambda^2 m^6}{\rho} + \frac{\rho}{3} \rp\,. 
\ee
The solution is
\begin{align}
\rho (r) = \sqrt{12 \Lambda ^2 m^6 \left[ \lp \frac{1-2 G M_\text{\tiny BH}/r_h}{1-2 G M_\text{\tiny BH}/r} \rp^{1/3}-1\right]+\rho_0^2 \lp \frac{1-2 G M_\text{\tiny BH}/r_h}{1-2 G M_\text{\tiny BH}/r} \rp^{1/3}}\,,
\label{rho rel 3}
\end{align}
where we have once again fixed the integration constant by requiring~$\rho (r_h) = \rho_0$. In the non-relativistic regime ($r,r_h \gg 2 G M_\BH$), this reduces to
\be
\label{rhoint3}
\rho (r) = \sqrt{8 \Lambda^2 m^6 G M_\BH \lp \frac{1}{r} - \frac{1}{r_h} \rp + \rho_0^2}\,,
\ee
which solves the non-relativistic condition of hydrostatic equilibrium (Eq.~\eqref{hydro 3}). In particular, at short distances the profile becomes a power-law $\rho \sim 1/\sqrt{r}$:
\be
\label{intsim3}
\rho(r) \simeq \sqrt{\frac{8 \Lambda^2 m^6 G M_\BH}{r}}\,,
\ee
which is a milder growth than the~$1/r$ profile (Eq.~\eqref{intsim2}) for the case~$P_2 \propto \rho^2$.

As discussed above, this assumption of superfluidity is only valid for large enough distances such that DM is sub-critical. 
Substituting the relativistic density profile of Eq.~\eqref{rho rel 3}, once again with $r_h \gg 2 G M_\BH$ and neglecting the subleading contribution proportional to $\rho_0$,
the criterion for degeneracy (Eq.~\eqref{v deg}) gives 
\be
v^2 < \frac{(12)^{1/3} 6 \pi}{\zeta(3/2)} \lp \frac{\Lambda}{m} \rp^{2/3} \llp \frac{1}{(1-2 v^2)^{1/3}} -1\rrp^{1/3}\,.
\ee
It is easy to see that this can only be satisfied in the regime $v^2 \ll 1$, {\it i.e.}, as long as
\be
v < v_\text{\tiny deg} \simeq  0.3 \lp \frac{m}{\rm eV} \rp^{-1/2}
\lp \frac{\Lambda}{\rm meV} \rp^{1/2}\,.
\label{v deg 3body}
\ee
Thus the DM particles cease to be degenerate around the same distance as they become relativistic. The corresponding radius at which the degeneracy condition is broken is estimated to be
\be
r_\text{\tiny deg} \simeq \frac{G M_\BH \zeta(3/2)}{8 (3 \pi)^{3/2}} \frac{m}{\Lambda}
\simeq 5.4 \cdot 10^{-6} \, {\rm kpc} 
\lp \frac{M_\BH}{10^{10} M_\odot} \rp \lp \frac{m}{\rm eV} \rp
\lp \frac{\Lambda}{\rm meV} \rp^{-1}\,.
\label{r deg 3body}
\ee
Comparing this value with Eq.~\eqref{Schwarzschild} shows that the degeneracy condition is violated before the BH horizon is reached. 
For~$r < r_\text{\tiny deg}$, the fluid equation of state is more aptly described by the ideal gas law,~$P \propto \rho v^2$. This regime will be discussed below.

Note that the fluid sound speed (Eq.~\eqref{cs3}) at the degeneracy radius is given by
\be
c_s^2 (r_\text{\tiny deg}) = \frac{\rho^2 (r_\text{\tiny deg})}{4\Lambda^2m^6} \simeq 0.2 \frac{\Lambda}{{\rm meV}}\frac{{\rm eV}}{m}  \,.
\ee
This implies that the non-relativistic equation of state~$P\sim \rho^3$ assumed above should receive significant corrections before reaching~$r_\text{\tiny deg}$. As in the two-body case, we ignore this issue for simplicity.

\subsection{Inner region (three-body case)}
\label{inner region}

We have seen that, while the two-body interacting superfluid remains degenerate all the way to the accretion radius~$r = 4G M_\BH$, this is not so in the three-body case. For the latter, degeneracy breaks down at~$r_\text{\tiny deg}$, which is larger than~$4GM_\BH$.  

Our goal is to derive the DM density and velocity profiles in this innermost region,~$4GM_\BH< r < r_\text{\tiny deg}$. We follow the results outlined in Ref.~\cite{Shapiro:2014oha}, where the interested reader can find additional details. Let us stress that, for simplicity, we neglect the role that gravitational scattering off stars plays in establishing the steady-state distribution of the dark matter around the central object, see Ref.~\cite{Shapiro:2022prq} for a recent analysis.

As mentioned already, although the DM component is no longer degenerate in this region, it nevertheless remains thermal, thanks to the large number of interactions around the central BH. Therefore its equation of state can be approximated as an ideal gas,
\be
P = n k_{\rm B} T = \frac{\rho v^2}{3}\,.
\label{ideal}
\ee
Within full general relativity, the density and temperature profile of the dark matter particles around the central BH are determined by the condition of hydrostatic equilibrium, given by Eq.~\eqref{hydro rel}, together with a heat equation within the gravothermal fluid approximation~\cite{Balberg:2002ue, 1980MNRAS.191..483L, Shapiro:2014oha, Koda:2011yb}:
\be
q_r = - \frac{\kappa}{\sqrt{1-\frac{2GM_\BH}{r}}} \,\frac{{\rm d}}{{\rm d}r} \left(\sqrt{1-\frac{2GM_\BH}{r}} \,T\right) = -\kappa \left( \frac{\d T}{\d r} + 
    \frac{T}{1-2GM_\BH/r}\frac{G M_\BH}{r^2} \right)\,.
\label{heat 1}
\ee
where~$q_r$ is the radial component of the heat flux. For a virialized gas at rest in a stationary, spherically-symmetric gravitational field,~$q_r$ is the only non-zero component of the heat flux four-vector~\cite{17b23f5522334b1ca8bb575ecaf5c01e}. Equation~\eqref{heat 1} is a relativistic Fourier's law, which states that the heat flux is proportional to the temperature gradient, with proportionality constant given by the conductivity~$\kappa$.
 
Assuming a steady-state solution~\cite{Bahcall:1976aa}, one can show that the DM cluster is virialized and at rest on a dynamical time scale, such that the mean fluid velocity is everywhere negligible, resulting into a constant total radiated heat:
\begin{equation}
r^2 q_r = \frac{D}{\sqrt{1-\frac{2GM_\BH}{r}}}\,,
\end{equation}
where~$D$ is a constant. Using this, Eq.~\eqref{heat 1} implies
\be
\frac{\d T}{\d r} = \frac{D}{\kappa r^2 (1-2 G M_\BH/r)^{3/2}} - \frac{T}{1-2 G M_\BH/r} \frac{G M_\BH}{r^2}\,.
\label{heat 2}
\ee
The only quantity left to specify is the thermal conductivity $\kappa$. 
In the non-relativistic limit, using equipartition~$k_{\rm B}T = \frac{1}{3}m v^2$, one can show that it is given in terms of the scattering cross section~$\sigma$ and DM particle mass~$m$ by~\cite{1980MNRAS.191..483L,Gnedin:2000ea,Koda:2011yb}\footnote{The expression for the thermal conductivity is obtained by matching the relativistic heat flux equation to the Newtonian result of kinetic theory, as done in Ref.~\cite{Shapiro:2014oha}.}
\be
\kappa = \frac{\sqrt{3}}{2} k_{\rm B} v \lp \frac{\sigma}{B} + \frac{m^2}{A C \sigma \rho^2 H^2} \rp^{-1}
\label{kappa def}
\ee
where $C \approx \frac{290}{385}$ is a constant determined by N-body simulations~\cite{Koda:2011yb},~$B = \frac{25 \sqrt{\pi}}{32}$ is found perturbatively in Chapman-Enskog theory~\cite{1981phki.book.....L}, and $A = \sqrt{\frac{16}{\pi}}$ for hard-sphere interactions. For gravitationally bound particles, the gravitational scale height~$H$ is the minimum of
the radius~$r$ and the Jeans scale:
\be
H = {\rm min} \left(r, \sqrt{\frac{v^2}{4 \pi G \rho}}\right)\,.
\ee
Which of the two contributions in Eq.~\eqref{kappa def} dominates the thermal conductivity depends on the ratio~$\frac{\ell}{H}$, where~$\ell= \frac{m}{\rho\sigma}$ is the mean free path (MFP). The first contribution dominates in the short MFP regime,~$\ell \ll H$, while the second dominates in the long MFP regime,~$\ell \gg H$.

In our case, it turns out that~$H \simeq r$. To see this, we can compare~$r$ and~$\sqrt{v^2/4 \pi G \rho}$ at the degeneracy radius. 
Using Eqs.~\eqref{intsim3},~\eqref{v deg 3body} and~\eqref{r deg 3body}, we obtain
\begin{align}
 \sqrt{\frac{v^2_\text{\tiny deg}}{4 \pi G \rho (r_\text{\tiny deg})}}\simeq 4 \times 10^7 \, r_\text{\tiny deg}\lp \frac{M_\BH}{10^{10} M_\odot} \rp^{-1} \lp \frac{m}{\rm eV} \rp^{-11/4} \lp \frac{\Lambda}{\rm meV} \rp^{3/4}  \gg r_\text{\tiny deg} \,.
\end{align}
As we will see, this hierarchy is maintained when one considers the radial evolution of the velocity and density profile for distances $r \lesssim r_\text{\tiny deg}$.
Therefore we henceforth set
\be
H = r\,.
\ee
Furthermore, it can be similarly argued that we are in the long MFP regime. Indeed, at the degeneracy radius,
\be
\frac{\ell}{r_\text{\tiny deg}} = \frac{m}{\sigma \rho(r_\text{\tiny deg}) r_\text{\tiny deg}} \simeq 3 \times 10^5 \left(\frac{\sigma/m}{{\rm cm}^2/{\rm g}}\right)^{-1}\lp \frac{M_\BH}{10^{10} M_\odot} \rp^{-1} \lp \frac{m}{\rm eV} \rp^{-7/2} \lp \frac{\Lambda}{\rm meV} \rp^{-1/2}\,.
\ee
We will see that this ratio increases as~$r$ decreases. It follows that
\be
\kappa \simeq \frac{\sqrt{3}}{2} A C k_{\rm B} \frac{\sigma\rho^2 r^2v}{m^2} \,.
\ee

In the relativistic regime, temperature and velocity dispersion are related by $k_{\rm B} T = \frac{1}{3}\gamma m v^2$, with~$\gamma = \frac{1}{\sqrt{1-v^2}}$.
Using this, together with Eq.~\eqref{ideal}, the condition of hydrostatic equilibrium (Eq.~\eqref{hydro rel}) and the heat equation~\eqref{heat 2} can be cast
as equations for density and velocity: 
\begin{align}
\frac{1}{\rho}\frac{\d (\rho v^2)}{\d r} &= - \frac{3 + v^2}{1-2 G M_\BH/r} \frac{G M_\BH}{r^2}\,; \nonumber \\
\frac{\d (\gamma v^2)}{\d r} &= \frac{3 k_{\rm B} D}{m \kappa r^2 (1-2 G M_\BH/r)^{3/2}} - \frac{\gamma v^2}{1-2 G M_\BH/r} \frac{G M_\BH }{r^2}\,. 
\label{rho v eqns 1}
\end{align}
Following Ref.~\cite{Shapiro:2014oha}, it is convenient to define dimensionless Newtonian quantities
\be
\rho_{\rm N} \equiv \frac{\rho}{\gamma \rho (r_\text{\tiny deg})} = \frac{mn}{\rho (r_\text{\tiny deg})}\,;\qquad v_{\rm N}^2 \equiv \frac{\gamma v^2}{v_\text{\tiny deg}^2}\,, 
\ee
normalized at the degeneracy radius. Defining also a dimensionless radius~$\tilde r = r/r_\text{\tiny deg}$, Eqs.~\eqref{rho v eqns 1} become
\begin{align}
\frac{\d \rho_{\rm N}}{\d {\tilde r}} &= - \frac{3\gamma}{v_{\rm N}^2 (1-2 v_\text{\tiny deg}^2/{\tilde r})} \frac{\rho_{\rm N}}{{\tilde r}^2} -\frac{2}{v_{\rm N}^3 (1-2 v_\text{\tiny deg}^2/{\tilde r})^{3/2}} \frac{\tilde D}{\rho_{\rm N} \gamma^{3/2} {\tilde r}^4}\,; \nonumber \\
\frac{\d v_{\rm N}}{\d {\tilde r}} &= \frac{1}{v_{\rm N}^2 (1-2 v_\text{\tiny deg}^2/{\tilde r})^{3/2}}\frac{\tilde D}{\rho_{\rm N}^2 \gamma^{3/2} {\tilde r}^4} - \frac{v_{\rm N}}{2(1-2 v_\text{\tiny deg}^2/{\tilde r})} \frac{v_\text{\tiny deg}^2 }{r^2}\,,
\label{dimensionless eqns}
\end{align}
where
\be
\gamma = \frac{v_{\rm N}^2 v_\text{\tiny deg}^2}{2} + \sqrt{1+\frac{v_{\rm N}^4 v_\text{\tiny deg}^4}{4}}\,; \qquad
\tilde D = \frac{\sqrt{3}m D}{A C \sigma r_\text{\tiny deg}^3 \rho (r_\text{\tiny deg})^2 v_\text{\tiny deg}^3}\,.
\ee
The constant $\tilde{D}$ can be estimated by requiring that 
particles deep inside the gravitational potential well of the black hole,
but outside its horizon, plunge directly into the black hole as accretion processes. This occurs at the radius~\cite{Sadeghian:2013laa,Shapiro:2014oha}
\begin{equation}
r_\text{\tiny mb} = 4 G M_\BH\,.
\end{equation}
Such a radius corresponds to marginally bound circular orbits in a Schwarzschild geometry, with energy $M_\BH$ and angular momentum per unit mass $4 G M_\BH$. A particle with larger angular momentum and smaller energy has an inner turning point at $r \gtrsim 4 G M_\BH$. Thus any particle that reaches it, would be necessarily captured by the BH.\footnote{We notice that this radius corresponds to the Bondi radius $r_\text{\tiny B} = G M_\BH/v_\text{\tiny rel}^2$ of a BH accreting particles with relativistic velocities $v_\text{\tiny rel} \approx 1/2$~\cite{1983bhwd.book.....S}. This quantity describes the region of space around the BH where accretion is relevant, and we therefore expect that already at distances $r \gtrsim r_\text{\tiny mb}$ the assumption of static equilibrium may be violated since accretion would be relevant. We leave this refinement to future work.}
The net result is that we can set an inner boundary at $r=r_\text{\tiny mb}$ where the DM density plummets and can be approximated to be equal to zero.
The value of $\tilde{D}$ is determined numerically by finding the point at which the energy density reaches its maximum ($\rho' = 0$ and $\rho'' <0$) and then drops off, giving $\tilde{D} \simeq -1.37$ for the values of the model parameters chosen above.

In the next Section we will show the result of numerically integrating Eqs.~\eqref{dimensionless eqns}. It turns out that the density profile in the inner region ($r_\text{\tiny mb} < r < r_\text{\tiny deg}$) is well-approximated by a power-law~\cite{Shapiro:2014oha}
\be
\rho (r) \simeq \rho (r_\text{\tiny deg}) \lp \frac{r}{r_\text{\tiny deg}} \rp^{-3/4}, \qquad r \lesssim r_\text{\tiny deg}\,,
\label{inner power-law}
\ee
until it reaches the inner boundary where $\rho (r_\text{\tiny mb}) \to 0$. This power-law is the same as in self-interacting DM~\cite{Shapiro:2014oha}.
Using these results, it is easy to confirm that the approximations~$H \simeq r$ and~$\ell \gg r$ are justified over the entire inner region.

\section{Dark matter profile and mass function}

In this Section we can finally review the full shape of the superfluid DM density profile and the corresponding mass function. Following the structure of the previous sections,
we discuss the results for both equations of state of interest. For this purpose, it will be instructive to compare the superfluid DM results with those for collisionless (CDM) and self-interacting DM (SIDM). Their density profiles are approximately power-law,
\be
\rho(r) \simeq \rho_0 \left(\frac{r}{r_h}\right)^{-\beta}\,, \qquad (r_\text{\tiny mb} \lesssim r \lesssim r_h)\,,
\ee
with~$\beta = 9/4$ (CDM) and $3/4$ (SIDM, for velocity-independent cross-section). The corresponding mass function is
\be
M(r) =  4\pi \int_{r_\text{\tiny mb}}^r \rho (r') r'^2 \d r' = \frac{4\pi}{3-\beta} \rho_0 r_h^3 \left[ \left(\frac{r}{r_h}\right)^{3-\beta} - \left(\frac{r_\text{\tiny mb}}{r_h}\right)^{3-\beta}\right]\,.
\label{M(r) CDM/SIDM}
\ee

\begin{figure}[t!]
	\centering
	\includegraphics[width=0.49\textwidth]{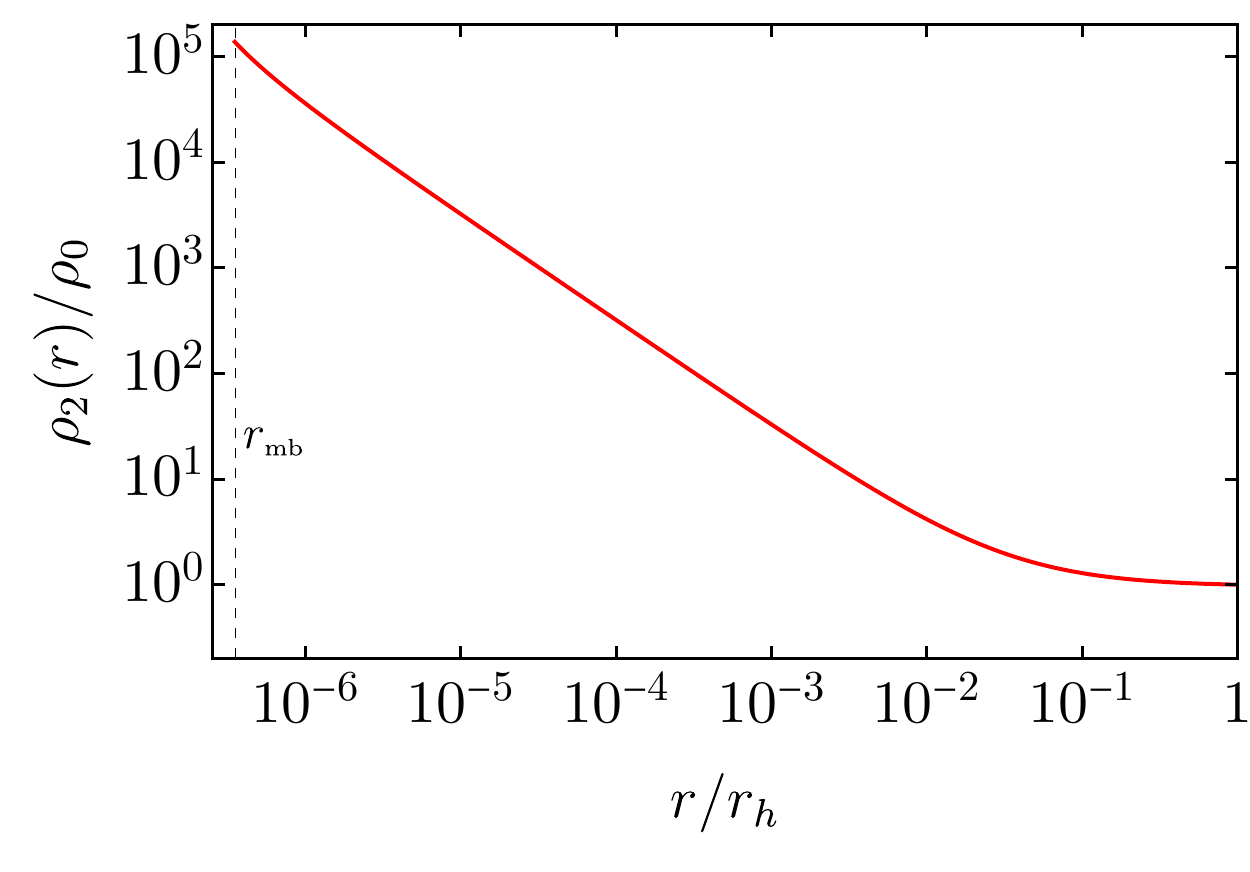}
    \includegraphics[width=0.49\textwidth]{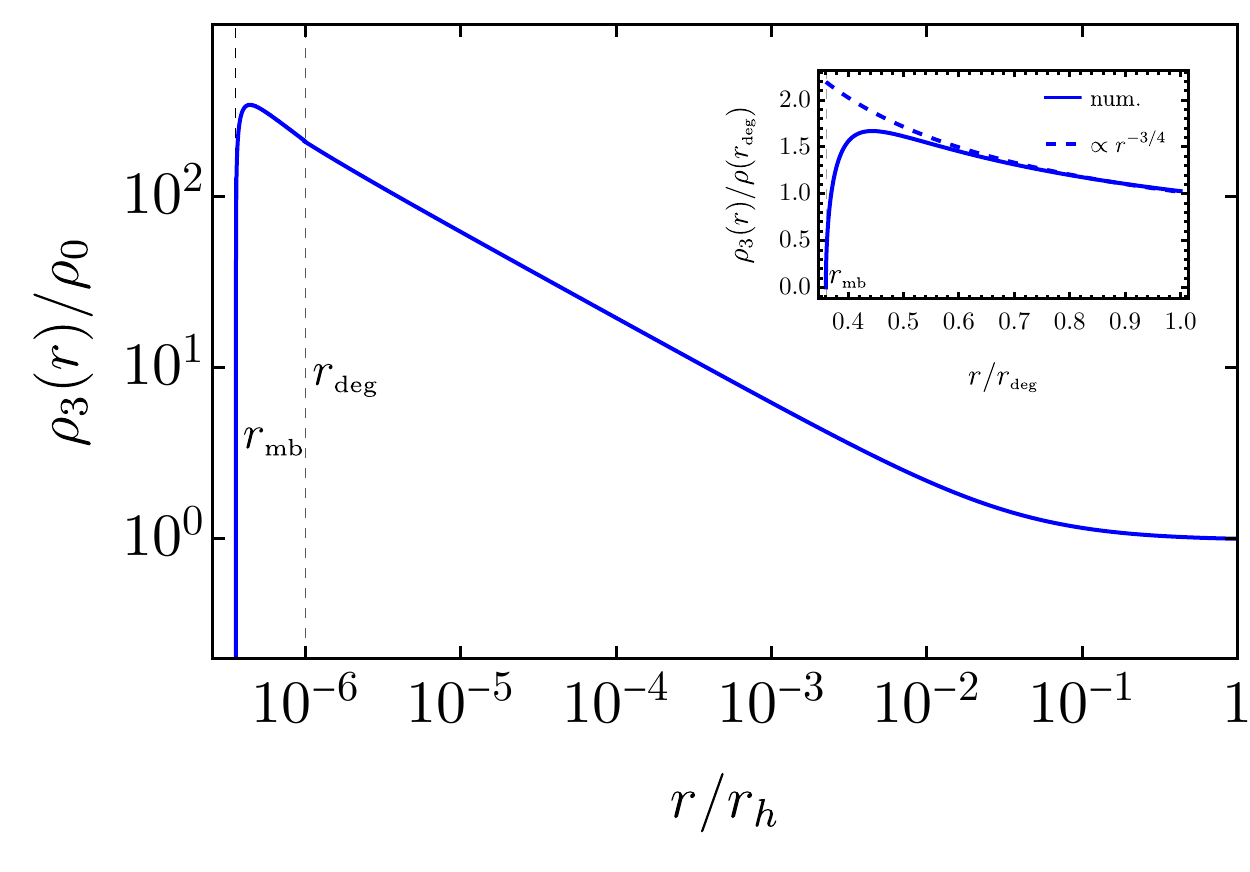}
	\caption{Density profiles of the superfluid DM around a supermassive black hole with mass~$M_\BH = 10^{10} M_\odot$, for the equations of state $P \propto \rho^2$ (Left, with~$m = \mu {\rm eV}$ and $a = 10^{-11} \, {\rm fm}$) and $P \propto \rho^3$ (Right, with~$m = {\rm eV}$ and $\Lambda = {\rm meV}$). The vertical dashed lines denote the position of the accretion radius $r_\text{\tiny mb}$, and the degeneracy radius $r_\text{\tiny deg}$. The inset in the Right Panel shows the comparison between the density profile in the inner region found numerically (solid line) and the power-law fit~$\rho \propto r^{-3/4}$ (dashed line).}
	\label{profile}
\end{figure}

\subsection{Two-body interacting superfluid}

For the two-body equation of state~$P\propto \rho^2$, the superfluid DM density profile found in the previous Sections (Eqs.~\eqref{rho core2} and~\eqref{rho rel}) is given by
\begin{align}
\rho_2(r) =
\begin{cases}
 \vspace{.2cm}
\frac{m^3}{2\pi a} \left[\left(\frac{1-2GM_\BH/r_h}{1-2GM_\BH/r}\right)^{1/4}-1\right] + \left(\frac{1-2GM_\BH/r_h}{1-2GM_\BH/r}\right)^{1/4}\rho_0 \qquad~~  r_\text{\tiny mb} \lesssim r \lesssim r_h\,;\\
\vspace{.2cm}
\rho_0 \frac{{\rm sin}\llp \sqrt{\frac{G m^3}{a}}(r-r_h)\rrp}{\sqrt{\frac{G m^3}{a}}(r-r_h)}
\qquad~~~~~~~~~~~~~~~~~~~~~~~~~~~~~~~~~~~~~~~~~~~~r_h \lesssim r \lesssim r_h + R_2\,.
\end{cases}
\end{align}
As discussed in Sec.~\ref{interm reg}, for the values of the scattering length~$a$ we are interested in, the DM remains degenerate all the way to the accretion radius~$r_\text{\tiny mb}$. 
Therefore, at the level of our approximations, we cannot resolve the behavior of the density profile in the vicinity of~$r_\text{\tiny mb}$, where~$\rho(r)$ should turn over and plummet.

Figure~\ref{profile} (Left Panel) shows the corresponding density profile around a~$10^{10} M_\odot$ supermassive black hole, with fiducial parameter values~$m = \mu {\rm eV}$ and~$a = 10^{-11} \, {\rm fm}$. As one can appreciate, the profile becomes increasingly steep as we approach the BH, until the accretion radius where~$\rho(r)$ is expected to drop. Our results show that the density can increase by orders of magnitude within the BH sphere of influence ($r < r_h$) with respect to the case where the BH is absent.

\begin{figure}[t!]
	\centering
	\includegraphics[width=0.49\textwidth]{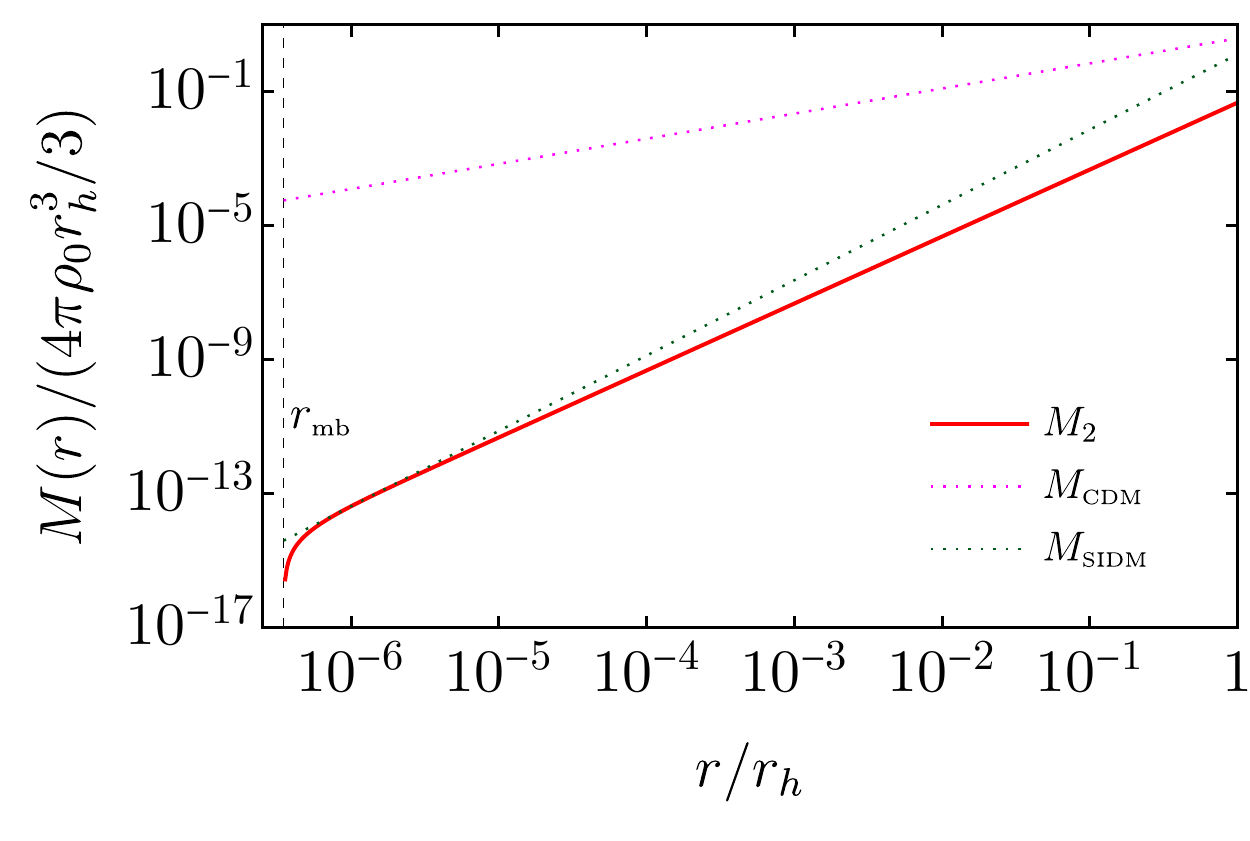}
    \includegraphics[width=0.49\textwidth]{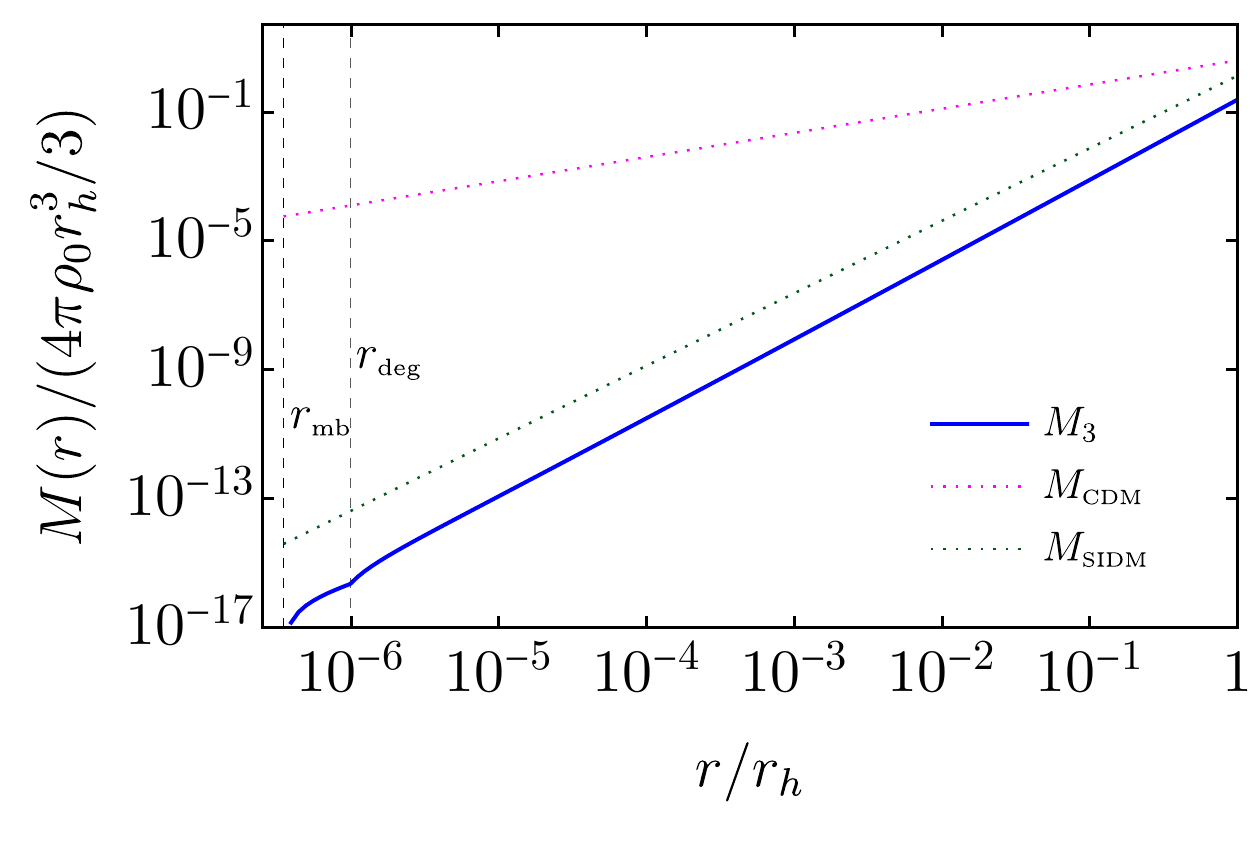}
	\caption{\it Superfluid DM halo mass profile around a supermassive black hole with mass $M_\BH = 10^{10} M_\odot$, assuming the equations of state $P \propto \rho^2$ (left) and $P \propto \rho^3$ (right). The dotted magenta and green lines indicate the mass function for the models of collisionless and self-interacting dark matter. We stress that we have fixed the central density $\rho_0$ to compute the mass function.}
	\label{haloprofile}
\end{figure}

Let us now evaluate the DM mass function,~$M_2(r) = 4\pi \int_{r_\text{\tiny mb}}^r \rho_2 (r') r'^2 \d r'$. 
This gives
\begin{align}
M_{2} (r) = 
\begin{cases}
 \vspace{.2cm}
\frac{m^3}{2a}G  M_\BH \left(r^2-r_\text{\tiny mb}^2 \right) \qquad~~~~~~~~~~~~~~~~~~~~~~~~~~~~~~~~~   r_\text{\tiny mb}  \lesssim r \lesssim r_h\,; \\
 \frac{m^3}{2a}G  M_\BH \left(r^2_h-r_\text{\tiny mb}^2\right) + \frac{4}{3} \pi \rho_0 \left(r^3 - r_h^3\right)  \qquad~~~~~~~~~~~~  r_h  \lesssim r \lesssim r_h + R_2 \,.
\end{cases}
\end{align}
To obtain a simpler analytical expression, we have used the approximate power-law profile~$\rho_2(r) \sim 1/r$ of Eq.~\eqref{intsim2} in the region~$r_\text{\tiny mb}  \lesssim r \lesssim r_h$, and assumed a constant profile~$\rho_2(r) \simeq \rho_0$ in the outer region~$r_h  \lesssim r \lesssim r_h + R_2$.   

Figure~\ref{haloprofile} (Left Panel) shows the DM halo mass function for the same parameter values as above. For comparison, we also plot the approximate halo mass functions for CDM and SIDM, given by Eq.~\eqref{M(r) CDM/SIDM}. Our results show that a two-body interacting superfluid is characterized by a different profile slope compared to CDM and SIDM. In particular, the presence of strong interactions within the superfluid model give rise to a shallower growth with respect to the other two cases, resulting in a less steep profile. Physically this is due to the classical pressure in the superfluid state.

\subsection{Three-body interacting superfluid}
\label{3body subsec}

For the three-body equation of state~$P\propto \rho^3$, the superfluid DM density profile is given analytically in the outer ($r_h \lesssim r \lesssim r_h + R_3$) and intermediate ($r_\text{\tiny deg} \lesssim r \lesssim r_h$) regions respectively by Eqs.~\eqref{rho core3} and~\eqref{rho rel 3}:
\begin{align}
\rho_3(r) =
\begin{cases}
 \vspace{.2cm}
\sqrt{12 \Lambda ^2 m^6 \left[ \lp \frac{1-2 G M_\text{\tiny BH}/r_h}{1-2 G M_\text{\tiny BH}/r} \rp^{1/3}-1\right]+\rho_0^2 \lp \frac{1-2 G M_\text{\tiny BH}/r_h}{1-2 G M_\text{\tiny BH}/r} \rp^{1/3}} \qquad  r_\text{\tiny deg} \lesssim r \lesssim r_h \,;\\
\vspace{.2cm}
\rho_0 \cos^{1/2}\left[\frac{\pi (r-r_h)}{2R_3}\right] \qquad~~~~~~~~~~~~~~~~~~~~~~~~~~~~~~~~~~~~~~~~~~~~~~~~ r_h \lesssim r \lesssim r_h + R_3\,.
\end{cases}
\end{align}
The density profile in the inner region ($r_\text{\tiny mb} < r < r_\text{\tiny deg}$) is obtained by numerically solving Eqs.~\eqref{dimensionless eqns} as described in Sec.~\ref{inner region}. 

In Fig.~\ref{profile} (Right Panel) we show the resulting density profile for fiducial parameter values~$m = {\rm eV}$ and $\Lambda = {\rm meV}$. As shown in the inset, the profile in the inner region is well-approximated by the power-law~$\rho(r) \sim r^{-3/4}$, given by Eq.~\eqref{inner power-law}, which is found to coincide with the SIDM behavior. Comparing the two panels, one can appreciate that the profile for~$P\propto \rho^3$ increases slightly less than in the $P \propto \rho^2$ case. This can be understood as follows. For the fiducial parameter values assumed in the two cases, the ratio of pressures is
\be
\frac{P_3}{P_2} = \frac{\rho \, m_2^3}{24 \pi a \Lambda^2m_3^6}  \simeq \lp \frac{\rho}{\rho_0} \rp 
 \lp \frac{\rho_0}{10^{-24}\, {\rm g/cm^3}} \rp
\lp \frac{m_2}{\rm \mu eV} \rp^{3} \lp \frac{a}{10^{-7} \, {\rm fm}} \rp^{-1} \lp \frac{\Lambda}{\rm meV} \rp^{-2} \lp \frac{m_3}{\rm eV} \rp^{-6}\,.
\label{P3overP2}
\ee 
In other words, as we go to smaller radii and the  density increases, the pressure in the three-body case is significantly larger, thereby allowing an overall growth in density which is milder in the three-body than in the two-body case.

Let us briefly comment on the velocity profile~$v(r)$. In the inner region ($r_\text{\tiny mb} < r < r_\text{\tiny deg}$), the velocity profile is determined by numerically integrating Eqs.~\eqref{dimensionless eqns}. The result is that~$v(r)$ grows steeply as we approach the accretion radius $r_\text{\tiny mb}$. The increase is much sharper than the~$1/\sqrt{r}$ scaling in the intermediate region. See Fig.~2 of Ref.~\cite{Shapiro:2014oha}.

The DM mass function in this case is given by
\begin{align}
M_3 (r) = 
\begin{cases}
 \vspace{.2cm}
\frac{32\sqrt{2}\pi}{9} \Lambda  m^3 \sqrt{G M_\BH} r_\text{\tiny deg}^{1/4} \left(r^{9/4} -r_\text{\tiny mb}^{9/4} \right) \qquad~~~~~~~~~~~~~~~~~~~~~~~~~~~~   r_\text{\tiny mb}  \lesssim r \lesssim r_\text{\tiny deg}\,; \\
 \vspace{.2cm}
16\sqrt{2}\pi \Lambda  m^3 \sqrt{G M_\BH} \Big\{\frac{2}{9}  r_\text{\tiny deg}^{1/4} \left(r^{9/4}_\text{\tiny deg} -r_\text{\tiny mb}^{9/4} \right) + \frac{1}{5} \left(r^{5/2}-r_\text{\tiny deg}^{5/2}\right)\Big\} ~~~ r_\text{\tiny deg}  \lesssim r \lesssim r_h\,; \\
 \vspace{.2cm}
\frac{16\sqrt{2} \pi}{5}   \Lambda  m^3 \sqrt{G M_\BH} \left(r_h^{5/2}-r_\text{\tiny deg}^{5/2}\right)+ \frac{4}{3} \pi \rho_0 \left(r^3 - r_h^3\right)  \qquad~~~~~~ r_h  \lesssim r \lesssim r_h + R_3 \,.
\end{cases}
\end{align}
Once again, to simplify the analytical expression, we have used the approximate power-law profile~$\rho_3(r) \sim 1/\sqrt{r}$ of Eq.~\eqref{intsim3} in the region~$r_\text{\tiny deg}  \lesssim r \lesssim r_h$, and assumed a constant profile~$\rho_3(r) \simeq \rho_0$ in the outer region~$r_h  \lesssim r \lesssim r_h + R_3$. Figure~\ref{haloprofile} (Right Panel) shows the DM halo mass function for the same parameter values as before, as well as CDM and SIDM mass functions (Eq.~\eqref{M(r) CDM/SIDM}) for comparison. Comparing the two panels, we notice that the halo mass for~$P\propto \rho^3$ assumes comparable values to the $P \propto \rho^2$ case.

\section{Conclusions}

The novel theory of DM superfluidity is able to reconcile the triumph of the $\Lambda$CDM model on cosmological scales while giving rise to a rich phenomenology on galactic scales. Within this model, the DM is represented by self-interacting particles which are generated out-of-equilibrium and remain decoupled from baryons throughout the history of the universe. These particles are able to thermalize and form a superfluid in galaxies, with critical temperature of order $\sim$mK, if their mass is sufficiently light and their self-interactions sufficiently strong.

At the center of these galaxies, large and massive black holes are expected to be present~\cite{Kormendy:2013dxa,vandenBosch:2016ild}, and they should currently contribute to about $10^{-5}$ of the DM in the universe~\cite{Yu:2002sq,Shankar:2007zg}. We therefore expect that the presence of these massive objects will modify the density profile of the surrounding DM fluid, with different predictions depending on the nature of the dark matter one considers.

In this work we computed the density profile of superfluid DM around central SMBHs, for different superfluid equations of state,~$P \propto \rho^2$ and $P \propto \rho^3$, corresponding to predominantly two-body and three-body interactions, respectively. We found that, depending on the distance from the central object, the DM density is characterized by different functional behavior, providing striking distinguishing features from standard predictions for collisionless DM~\cite{Bertschinger:1985pd}.

One possible way to reveal the underlying properties of DM, and eventually its superfluid nature, could be through gravitational lensing observations~\cite{Tambalo:2022wlm, Qiao:2022nic}. In particular, it is expected that the gravitational lensing of central black holes is influenced by the surrounding DM, which would modify for example the photon sphere and black hole shadow, thus providing a novel way to probe DM properties. Moreover, continuous gravitational waves emitted by isolated neutron stars and lensed by SMBHs like Sgr A$^*$ might be observable with future gravitational wave detectors. As such they can provide a new probe of the matter distribution in the galactic center~\cite{Savastano:2022jjv}.
Furthermore, it might be possible to probe this kind of DM profiles if we can assess that a binary merged near a SMBH, for instance via the detection of the gravitational spin Hall effect, see Ref.~\cite{Oancea:2022szu} for further details.

To improve on our analysis, it would be important to include the contribution of realistic baryon distributions, which are expected to modify the DM profile at distances larger than the BH sphere of influence, especially in the matching to the NFW envelope, as well as the role that gravitational scattering off stars plays in establishing the steady-state distribution of DM around the central object. Finally, the inclusion of an AGN disk around the central BH and the corresponding emission of jets would probably modify the density profile in the region close to the BH.
We leave these further improvements to future work.

\subsubsection*{Acknowledgments}
We thank Lasha Berezhiani, Giordano Cintia and Stuart Shapiro for illuminating comments and feedback on the draft. We are also grateful to Miguel Zumalacarregui for interesting discussions. We thank an anonymous referee for valuable comments and suggestions.
V.DL. is supported by funds provided by the Center for Particle Cosmology at the University of Pennsylvania. 
The work of J.K. is supported in part by the DOE (HEP) Award DE-SC0013528.

\bibliographystyle{JHEP}
\bibliography{draft.bib}

\providecommand{\href}[2]{#2}\begingroup\raggedright\begin{thebibliography}{10}

\bibitem{Bullock:2017xww}
J.S.~Bullock and M.~Boylan-Kolchin, \emph{{Small-Scale Challenges to the
  $\Lambda$CDM Paradigm}},
  \href{https://doi.org/10.1146/annurev-astro-091916-055313}{\emph{Ann. Rev.
  Astron. Astrophys.} {\bfseries 55} (2017) 343}
  [\href{https://arxiv.org/abs/1707.04256}{{\ttfamily 1707.04256}}].

\bibitem{Navarro:1995iw}
J.F.~Navarro, C.S.~Frenk and S.D.M.~White, \emph{{The Structure of cold dark
  matter halos}}, \href{https://doi.org/10.1086/177173}{\emph{Astrophys. J.}
  {\bfseries 462} (1996) 563}
  [\href{https://arxiv.org/abs/astro-ph/9508025}{{\ttfamily
  astro-ph/9508025}}].

\bibitem{Oman:2015xda}
K.A.~Oman et~al., \emph{{The unexpected diversity of dwarf galaxy rotation
  curves}}, \href{https://doi.org/10.1093/mnras/stv1504}{\emph{Mon. Not. Roy.
  Astron. Soc.} {\bfseries 452} (2015) 3650}
  [\href{https://arxiv.org/abs/1504.01437}{{\ttfamily 1504.01437}}].

\bibitem{McGaugh:2016leg}
S.~McGaugh, F.~Lelli and J.~Schombert, \emph{{Radial Acceleration Relation in
  Rotationally Supported Galaxies}},
  \href{https://doi.org/10.1103/PhysRevLett.117.201101}{\emph{Phys. Rev. Lett.}
  {\bfseries 117} (2016) 201101}
  [\href{https://arxiv.org/abs/1609.05917}{{\ttfamily 1609.05917}}].

\bibitem{Lelli:2016cui}
F.~Lelli, S.S.~McGaugh, J.M.~Schombert and M.S.~Pawlowski, \emph{{One Law to
  Rule Them All: The Radial Acceleration Relation of Galaxies}},
  \href{https://doi.org/10.3847/1538-4357/836/2/152}{\emph{Astrophys. J.}
  {\bfseries 836} (2017) 152}
  [\href{https://arxiv.org/abs/1610.08981}{{\ttfamily 1610.08981}}].

\bibitem{McGaugh:2000sr}
S.S.~McGaugh, J.M.~Schombert, G.D.~Bothun and W.J.G.~de~Blok, \emph{{The
  Baryonic Tully-Fisher relation}},
  \href{https://doi.org/10.1086/312628}{\emph{Astrophys. J. Lett.} {\bfseries
  533} (2000) L99} [\href{https://arxiv.org/abs/astro-ph/0003001}{{\ttfamily
  astro-ph/0003001}}].

\bibitem{McGaugh:2011ac}
S.~McGaugh, \emph{{The Baryonic Tully-Fisher Relation of Gas Rich Galaxies as a
  Test of LCDM and MOND}},
  \href{https://doi.org/10.1088/0004-6256/143/2/40}{\emph{Astron. J.}
  {\bfseries 143} (2012) 40} [\href{https://arxiv.org/abs/1107.2934}{{\ttfamily
  1107.2934}}].

\bibitem{Famaey:2011kh}
B.~Famaey and S.~McGaugh, \emph{{Modified Newtonian Dynamics (MOND):
  Observational Phenomenology and Relativistic Extensions}},
  \href{https://doi.org/10.12942/lrr-2012-10}{\emph{Living Rev. Rel.}
  {\bfseries 15} (2012) 10} [\href{https://arxiv.org/abs/1112.3960}{{\ttfamily
  1112.3960}}].

\bibitem{Papastergis:2016jqv}
E.~Papastergis, E.A.K.~Adams and J.M.~van~der Hulst, \emph{{An accurate
  measurement of the baryonic Tully-Fisher relation with heavily gas-dominated
  ALFALFA galaxies}},
  \href{https://doi.org/10.1051/0004-6361/201628410}{\emph{Astron. Astrophys.}
  {\bfseries 593} (2016) A39}
  [\href{https://arxiv.org/abs/1602.09087}{{\ttfamily 1602.09087}}].

\bibitem{Lelli:2015wst}
F.~Lelli, S.S.~McGaugh and J.M.~Schombert, \emph{{The Small Scatter of the
  Baryonic Tully\textendash{}fisher Relation}},
  \href{https://doi.org/10.3847/2041-8205/816/1/L14}{\emph{Astrophys. J. Lett.}
  {\bfseries 816} (2016) L14}
  [\href{https://arxiv.org/abs/1512.04543}{{\ttfamily 1512.04543}}].

\bibitem{DiCintio:2015eeq}
A.~Di~Cintio and F.~Lelli, \emph{{The mass discrepancy acceleration relation in
  a $\Lambda$CDM context}},
  \href{https://doi.org/10.1093/mnrasl/slv185}{\emph{Mon. Not. Roy. Astron.
  Soc.} {\bfseries 456} (2016) L127}
  [\href{https://arxiv.org/abs/1511.06616}{{\ttfamily 1511.06616}}].

\bibitem{Desmond:2016azy}
H.~Desmond, \emph{{A statistical investigation of the mass
  discrepancy\textendash{}acceleration relation}},
  \href{https://doi.org/10.1093/mnras/stw2571}{\emph{Mon. Not. Roy. Astron.
  Soc.} {\bfseries 464} (2017) 4160}
  [\href{https://arxiv.org/abs/1607.01800}{{\ttfamily 1607.01800}}].

\bibitem{Famaey:2017xou}
B.~Famaey, J.~Khoury and R.~Penco, \emph{{Emergence of the mass
  discrepancy-acceleration relation from dark matter-baryon interactions}},
  \href{https://doi.org/10.1088/1475-7516/2018/03/038}{\emph{JCAP} {\bfseries
  03} (2018) 038} [\href{https://arxiv.org/abs/1712.01316}{{\ttfamily
  1712.01316}}].

\bibitem{Famaey:2019baq}
B.~Famaey, J.~Khoury, R.~Penco and A.~Sharma, \emph{{Baryon-Interacting Dark
  Matter: heating dark matter and the emergence of galaxy scaling relations}},
  \href{https://doi.org/10.1088/1475-7516/2020/06/025}{\emph{JCAP} {\bfseries
  06} (2020) 025} [\href{https://arxiv.org/abs/1912.07626}{{\ttfamily
  1912.07626}}].

\bibitem{Spergel:1999mh}
D.N.~Spergel and P.J.~Steinhardt, \emph{{Observational evidence for
  selfinteracting cold dark matter}},
  \href{https://doi.org/10.1103/PhysRevLett.84.3760}{\emph{Phys. Rev. Lett.}
  {\bfseries 84} (2000) 3760}
  [\href{https://arxiv.org/abs/astro-ph/9909386}{{\ttfamily
  astro-ph/9909386}}].

\bibitem{Kaplinghat:2015aga}
M.~Kaplinghat, S.~Tulin and H.-B.~Yu, \emph{{Dark Matter Halos as Particle
  Colliders: Unified Solution to Small-Scale Structure Puzzles from Dwarfs to
  Clusters}}, \href{https://doi.org/10.1103/PhysRevLett.116.041302}{\emph{Phys.
  Rev. Lett.} {\bfseries 116} (2016) 041302}
  [\href{https://arxiv.org/abs/1508.03339}{{\ttfamily 1508.03339}}].

\bibitem{Hu:2000ke}
W.~Hu, R.~Barkana and A.~Gruzinov, \emph{{Cold and fuzzy dark matter}},
  \href{https://doi.org/10.1103/PhysRevLett.85.1158}{\emph{Phys. Rev. Lett.}
  {\bfseries 85} (2000) 1158}
  [\href{https://arxiv.org/abs/astro-ph/0003365}{{\ttfamily
  astro-ph/0003365}}].

\bibitem{Hui:2016ltb}
L.~Hui, J.P.~Ostriker, S.~Tremaine and E.~Witten, \emph{{Ultralight scalars as
  cosmological dark matter}},
  \href{https://doi.org/10.1103/PhysRevD.95.043541}{\emph{Phys. Rev. D}
  {\bfseries 95} (2017) 043541}
  [\href{https://arxiv.org/abs/1610.08297}{{\ttfamily 1610.08297}}].

\bibitem{Ferreira:2020fam}
E.G.M.~Ferreira, \emph{{Ultra-light dark matter}},
  \href{https://doi.org/10.1007/s00159-021-00135-6}{\emph{Astron. Astrophys.
  Rev.} {\bfseries 29} (2021) 7}
  [\href{https://arxiv.org/abs/2005.03254}{{\ttfamily 2005.03254}}].

\bibitem{Goodman:2000tg}
J.~Goodman, \emph{{Repulsive dark matter}},
  \href{https://doi.org/10.1016/S1384-1076(00)00015-4}{\emph{New Astron.}
  {\bfseries 5} (2000) 103}
  [\href{https://arxiv.org/abs/astro-ph/0003018}{{\ttfamily
  astro-ph/0003018}}].

\bibitem{Berezhiani:2015pia}
L.~Berezhiani and J.~Khoury, \emph{{Dark Matter Superfluidity and Galactic
  Dynamics}}, \href{https://doi.org/10.1016/j.physletb.2015.12.054}{\emph{Phys.
  Lett. B} {\bfseries 753} (2016) 639}
  [\href{https://arxiv.org/abs/1506.07877}{{\ttfamily 1506.07877}}].

\bibitem{Berezhiani:2015bqa}
L.~Berezhiani and J.~Khoury, \emph{{Theory of dark matter superfluidity}},
  \href{https://doi.org/10.1103/PhysRevD.92.103510}{\emph{Phys. Rev. D}
  {\bfseries 92} (2015) 103510}
  [\href{https://arxiv.org/abs/1507.01019}{{\ttfamily 1507.01019}}].

\bibitem{Berezhiani:2017tth}
L.~Berezhiani, B.~Famaey and J.~Khoury, \emph{{Phenomenological consequences of
  superfluid dark matter with baryon-phonon coupling}},
  \href{https://doi.org/10.1088/1475-7516/2018/09/021}{\emph{JCAP} {\bfseries
  09} (2018) 021} [\href{https://arxiv.org/abs/1711.05748}{{\ttfamily
  1711.05748}}].

\bibitem{Sharma:2018ydn}
A.~Sharma, J.~Khoury and T.~Lubensky, \emph{{The Equation of State of Dark
  Matter Superfluids}},
  \href{https://doi.org/10.1088/1475-7516/2019/05/054}{\emph{JCAP} {\bfseries
  05} (2019) 054} [\href{https://arxiv.org/abs/1809.08286}{{\ttfamily
  1809.08286}}].

\bibitem{Berezhiani:2018oxf}
L.~Berezhiani and J.~Khoury, \emph{{Emergent long-range interactions in
  Bose-Einstein Condensates}},
  \href{https://doi.org/10.1103/PhysRevD.99.076003}{\emph{Phys. Rev. D}
  {\bfseries 99} (2019) 076003}
  [\href{https://arxiv.org/abs/1812.09332}{{\ttfamily 1812.09332}}].

\bibitem{Slepian:2011ev}
Z.~Slepian and J.~Goodman, \emph{{Ruling Out Bosonic Repulsive Dark Matter in
  Thermal Equilibrium}},
  \href{https://doi.org/10.1111/j.1365-2966.2012.21901.x}{\emph{Mon. Not. Roy.
  Astron. Soc.} {\bfseries 427} (2012) 839}
  [\href{https://arxiv.org/abs/1109.3844}{{\ttfamily 1109.3844}}].

\bibitem{Markevitch:2003at}
M.~Markevitch, A.H.~Gonzalez, D.~Clowe, A.~Vikhlinin, L.~David, W.~Forman
  et~al., \emph{{Direct constraints on the dark matter self-interaction
  cross-section from the merging galaxy cluster 1E0657-56}},
  \href{https://doi.org/10.1086/383178}{\emph{Astrophys. J.} {\bfseries 606}
  (2004) 819} [\href{https://arxiv.org/abs/astro-ph/0309303}{{\ttfamily
  astro-ph/0309303}}].

\bibitem{Clowe:2003tk}
D.~Clowe, A.~Gonzalez and M.~Markevitch, \emph{{Weak lensing mass
  reconstruction of the interacting cluster 1E0657-558: Direct evidence for the
  existence of dark matter}},
  \href{https://doi.org/10.1086/381970}{\emph{Astrophys. J.} {\bfseries 604}
  (2004) 596} [\href{https://arxiv.org/abs/astro-ph/0312273}{{\ttfamily
  astro-ph/0312273}}].

\bibitem{Berezhiani:2021rjs}
L.~Berezhiani, G.~Cintia and M.~Warkentin, \emph{{Core fragmentation in
  simplest superfluid dark matter scenario}},
  \href{https://doi.org/10.1016/j.physletb.2021.136422}{\emph{Phys. Lett. B}
  {\bfseries 819} (2021) 136422}
  [\href{https://arxiv.org/abs/2101.08117}{{\ttfamily 2101.08117}}].

\bibitem{Sharma:2022jio}
A.~Sharma, G.~Kartvelishvili and J.~Khoury, \emph{{Finite temperature
  description of an interacting Bose gas}},
  \href{https://doi.org/10.1103/PhysRevD.106.045025}{\emph{Phys. Rev. D}
  {\bfseries 106} (2022) 045025}
  [\href{https://arxiv.org/abs/2204.02423}{{\ttfamily 2204.02423}}].

\bibitem{Berezhiani:2022buv}
L.~Berezhiani, G.~Cintia and J.~Khoury, \emph{{Thermalization, Fragmentation
  and Tidal Disruption: The Complex Galactic Dynamics of Dark Matter
  Superfluidity}},  \href{https://arxiv.org/abs/2212.10577}{{\ttfamily
  2212.10577}}.

\bibitem{10.1093/mnras/291.1.219}
R.~Genzel, A.~Eckart, T.~Ott and F.~Eisenhauer, \emph{{On the nature of the
  dark mass in the centre of the Milky Way}},
  \href{https://doi.org/10.1093/mnras/291.1.219}{\emph{Monthly Notices of the
  Royal Astronomical Society} {\bfseries 291} (1997) 219}
  [\href{https://arxiv.org/abs/https://academic.oup.com/mnras/article-pdf/291/1/219/4082303/291-1-219.pdf}{{\ttfamily
  https://academic.oup.com/mnras/article-pdf/291/1/219/4082303/291-1-219.pdf}}].

\bibitem{Eckart:1997em}
A.~Eckart and R.~Genzel, \emph{{Stellar proper motions in the central 0.1 PC of
  the galaxy}}, \href{https://doi.org/10.1093/mnras/284.3.576}{\emph{Mon. Not.
  Roy. Astron. Soc.} {\bfseries 284} (1997) 576}.

\bibitem{Ghez:2003qj}
A.M.~Ghez, S.~Salim, S.D.~Hornstein, A.~Tanner, M.~Morris, E.E.~Becklin et~al.,
  \emph{{Stellar orbits around the galactic center black hole}},
  \href{https://doi.org/10.1086/427175}{\emph{Astrophys. J.} {\bfseries 620}
  (2005) 744} [\href{https://arxiv.org/abs/astro-ph/0306130}{{\ttfamily
  astro-ph/0306130}}].

\bibitem{Richstone:1998ky}
D.~Richstone et~al., \emph{{Supermassive black holes and the evolution of
  galaxies}}, {\emph{Nature} {\bfseries 395} (1998) A14}
  [\href{https://arxiv.org/abs/astro-ph/9810378}{{\ttfamily
  astro-ph/9810378}}].

\bibitem{10.1007/978-94-011-4750-7_11}
L.C.~Ho, \emph{Supermassive black holes in galactic nuclei},  in
  \emph{Observational Evidence for Black Holes in the Universe},
  S.K.~Chakrabarti, ed., (Dordrecht), pp.~157--186, Springer Netherlands, 1999.

\bibitem{Madau_2001}
P.~Madau and M.J.~Rees, \emph{Massive black holes as population iii remnants},
  \href{https://doi.org/10.1086/319848}{\emph{The Astrophysical Journal}
  {\bfseries 551} (2001) L27}.

\bibitem{Shibata_2002}
M.~Shibata and S.L.~Shapiro, \emph{Collapse of a rotating supermassive star to
  a supermassive black hole: Fully relativistic simulations},
  \href{https://doi.org/10.1086/341516}{\emph{The Astrophysical Journal}
  {\bfseries 572} (2002) L39}.

\bibitem{Bean:2002kx}
R.~Bean and J.~Magueijo, \emph{{Could supermassive black holes be
  quintessential primordial black holes?}},
  \href{https://doi.org/10.1103/PhysRevD.66.063505}{\emph{Phys. Rev. D}
  {\bfseries 66} (2002) 063505}
  [\href{https://arxiv.org/abs/astro-ph/0204486}{{\ttfamily
  astro-ph/0204486}}].

\bibitem{Serpico:2020ehh}
P.D.~Serpico, V.~Poulin, D.~Inman and K.~Kohri, \emph{{Cosmic microwave
  background bounds on primordial black holes including dark matter halo
  accretion}},
  \href{https://doi.org/10.1103/PhysRevResearch.2.023204}{\emph{Phys. Rev.
  Res.} {\bfseries 2} (2020) 023204}
  [\href{https://arxiv.org/abs/2002.10771}{{\ttfamily 2002.10771}}].

\bibitem{DeLuca:2022bjs}
V.~De~Luca, G.~Franciolini and A.~Riotto, \emph{{Clusteringenesis: from Light
  to Heavy Primordial Black Holes}},
  \href{https://arxiv.org/abs/2210.14171}{{\ttfamily 2210.14171}}.

\bibitem{1972GReGr...3...63P}
P.J.E.~{Peebles}, \emph{{Gravitational collapse and related phenomena from an
  empirical point of view, or, black holes are where you find them.}},
  \href{https://doi.org/10.1007/BF00755923}{\emph{General Relativity and
  Gravitation} {\bfseries 3} (1972) 63}.

\bibitem{Gondolo:1999ef}
P.~Gondolo and J.~Silk, \emph{{Dark matter annihilation at the galactic
  center}}, \href{https://doi.org/10.1103/PhysRevLett.83.1719}{\emph{Phys. Rev.
  Lett.} {\bfseries 83} (1999) 1719}
  [\href{https://arxiv.org/abs/astro-ph/9906391}{{\ttfamily
  astro-ph/9906391}}].

\bibitem{Merritt:2003qk}
D.~Merritt, \emph{{Evolution of the dark matter distribution at the galactic
  center}}, \href{https://doi.org/10.1103/PhysRevLett.92.201304}{\emph{Phys.
  Rev. Lett.} {\bfseries 92} (2004) 201304}
  [\href{https://arxiv.org/abs/astro-ph/0311594}{{\ttfamily
  astro-ph/0311594}}].

\bibitem{Gnedin:2003rj}
O.Y.~Gnedin and J.R.~Primack, \emph{{Dark Matter Profile in the Galactic
  Center}}, \href{https://doi.org/10.1103/PhysRevLett.93.061302}{\emph{Phys.
  Rev. Lett.} {\bfseries 93} (2004) 061302}
  [\href{https://arxiv.org/abs/astro-ph/0308385}{{\ttfamily
  astro-ph/0308385}}].

\bibitem{Merritt:2006mt}
D.~Merritt, S.~Harfst and G.~Bertone, \emph{{Collisionally Regenerated Dark
  Matter Structures in Galactic Nuclei}},
  \href{https://doi.org/10.1103/PhysRevD.75.043517}{\emph{Phys. Rev. D}
  {\bfseries 75} (2007) 043517}
  [\href{https://arxiv.org/abs/astro-ph/0610425}{{\ttfamily
  astro-ph/0610425}}].

\bibitem{Shapiro:2022prq}
S.L.~Shapiro and D.C.~Heggie, \emph{{Effect of stars on the dark matter spike
  around a black hole: A tale of two treatments}},
  \href{https://doi.org/10.1103/PhysRevD.106.043018}{\emph{Phys. Rev. D}
  {\bfseries 106} (2022) 043018}
  [\href{https://arxiv.org/abs/2209.08105}{{\ttfamily 2209.08105}}].

\bibitem{Wanders:2014xia}
M.~Wanders, G.~Bertone, M.~Volonteri and C.~Weniger, \emph{{No WIMP Mini-Spikes
  in Dwarf Spheroidal Galaxies}},
  \href{https://doi.org/10.1088/1475-7516/2015/04/004}{\emph{JCAP} {\bfseries
  04} (2015) 004} [\href{https://arxiv.org/abs/1409.5797}{{\ttfamily
  1409.5797}}].

\bibitem{Vasiliev:2007vh}
E.~Vasiliev, \emph{{Dark matter annihilation near a black hole: Plateau vs.
  weak cusp}}, \href{https://doi.org/10.1103/PhysRevD.76.103532}{\emph{Phys.
  Rev. D} {\bfseries 76} (2007) 103532}
  [\href{https://arxiv.org/abs/0707.3334}{{\ttfamily 0707.3334}}].

\bibitem{Shapiro:2016ypb}
S.L.~Shapiro and J.~Shelton, \emph{{Weak annihilation cusp inside the dark
  matter spike about a black hole}},
  \href{https://doi.org/10.1103/PhysRevD.93.123510}{\emph{Phys. Rev. D}
  {\bfseries 93} (2016) 123510}
  [\href{https://arxiv.org/abs/1606.01248}{{\ttfamily 1606.01248}}].

\bibitem{Begelman:2006db}
M.C.~Begelman, M.~Volonteri and M.J.~Rees, \emph{{Formation of supermassive
  black holes by direct collapse in pregalactic halos}},
  \href{https://doi.org/10.1111/j.1365-2966.2006.10467.x}{\emph{Mon. Not. Roy.
  Astron. Soc.} {\bfseries 370} (2006) 289}
  [\href{https://arxiv.org/abs/astro-ph/0602363}{{\ttfamily
  astro-ph/0602363}}].

\bibitem{Ullio:2001fb}
P.~Ullio, H.~Zhao and M.~Kamionkowski, \emph{{A Dark matter spike at the
  galactic center?}},
  \href{https://doi.org/10.1103/PhysRevD.64.043504}{\emph{Phys. Rev. D}
  {\bfseries 64} (2001) 043504}
  [\href{https://arxiv.org/abs/astro-ph/0101481}{{\ttfamily
  astro-ph/0101481}}].

\bibitem{Fornasa:2007nr}
M.~Fornasa and G.~Bertone, \emph{{Black Holes as Dark Matter Annihilation
  Boosters}}, \href{https://doi.org/10.1142/S0218271808012747}{\emph{Int. J.
  Mod. Phys. D} {\bfseries 17} (2008) 1125}
  [\href{https://arxiv.org/abs/0711.3148}{{\ttfamily 0711.3148}}].

\bibitem{Shapiro:2014oha}
S.L.~Shapiro and V.~Paschalidis, \emph{{Self-interacting dark matter cusps
  around massive black holes}},
  \href{https://doi.org/10.1103/PhysRevD.89.023506}{\emph{Phys. Rev. D}
  {\bfseries 89} (2014) 023506}
  [\href{https://arxiv.org/abs/1402.0005}{{\ttfamily 1402.0005}}].

\bibitem{Feng:2021qkj}
W.-X.~Feng, A.~Parisi, C.-S.~Chen and F.-L.~Lin, \emph{{Self-interacting dark
  scalar spikes around black holes via relativistic Bondi accretion}},
  \href{https://doi.org/10.1088/1475-7516/2022/08/032}{\emph{JCAP} {\bfseries
  08} (2022) 032} [\href{https://arxiv.org/abs/2112.05160}{{\ttfamily
  2112.05160}}].

\bibitem{Chavanis:2019bnu}
P.-H.~Chavanis, \emph{{Mass-radius relation of self-gravitating Bose-Einstein
  condensates with a central black hole}},
  \href{https://doi.org/10.1140/epjp/i2019-12734-7}{\emph{Eur. Phys. J. Plus}
  {\bfseries 134} (2019) 352}
  [\href{https://arxiv.org/abs/1909.04709}{{\ttfamily 1909.04709}}].

\bibitem{Peacock:1999ye}
J.A.~Peacock, \emph{{Cosmological physics}} (1999).

\bibitem{Sikivie:2009qn}
P.~Sikivie and Q.~Yang, \emph{{Bose-Einstein Condensation of Dark Matter
  Axions}}, \href{https://doi.org/10.1103/PhysRevLett.103.111301}{\emph{Phys.
  Rev. Lett.} {\bfseries 103} (2009) 111301}
  [\href{https://arxiv.org/abs/0901.1106}{{\ttfamily 0901.1106}}].

\bibitem{Miralda-Escude:2000tvu}
J.~Miralda-Escude, \emph{{A test of the collisional dark matter hypothesis from
  cluster lensing}}, \href{https://doi.org/10.1086/324138}{\emph{Astrophys. J.}
  {\bfseries 564} (2002) 60}
  [\href{https://arxiv.org/abs/astro-ph/0002050}{{\ttfamily
  astro-ph/0002050}}].

\bibitem{Gnedin:2000ea}
O.Y.~Gnedin and J.P.~Ostriker, \emph{{Limits on collisional dark matter from
  elliptical galaxies in clusters}},
  \href{https://doi.org/10.1086/323211}{\emph{Astrophys. J.} {\bfseries 561}
  (2001) 61} [\href{https://arxiv.org/abs/astro-ph/0010436}{{\ttfamily
  astro-ph/0010436}}].

\bibitem{Randall:2008ppe}
S.W.~Randall, M.~Markevitch, D.~Clowe, A.H.~Gonzalez and M.~Bradac,
  \emph{{Constraints on the Self-Interaction Cross-Section of Dark Matter from
  Numerical Simulations of the Merging Galaxy Cluster 1E 0657-56}},
  \href{https://doi.org/10.1086/587859}{\emph{Astrophys. J.} {\bfseries 679}
  (2008) 1173} [\href{https://arxiv.org/abs/0704.0261}{{\ttfamily 0704.0261}}].

\bibitem{Navarro:1996gj}
J.F.~Navarro, C.S.~Frenk and S.D.M.~White, \emph{{A Universal density profile
  from hierarchical clustering}},
  \href{https://doi.org/10.1086/304888}{\emph{Astrophys. J.} {\bfseries 490}
  (1997) 493} [\href{https://arxiv.org/abs/astro-ph/9611107}{{\ttfamily
  astro-ph/9611107}}].

\bibitem{Dutton:2014xda}
A.A.~Dutton and A.V.~Macci\`o, \emph{{Cold dark matter haloes in the Planck
  era: evolution of structural parameters for Einasto and NFW profiles}},
  \href{https://doi.org/10.1093/mnras/stu742}{\emph{Mon. Not. Roy. Astron.
  Soc.} {\bfseries 441} (2014) 3359}
  [\href{https://arxiv.org/abs/1402.7073}{{\ttfamily 1402.7073}}].

\bibitem{Bernstein:1990kf}
J.~Bernstein and S.~Dodelson, \emph{{Relativistic Bose gas}},
  \href{https://doi.org/10.1103/PhysRevLett.66.683}{\emph{Phys. Rev. Lett.}
  {\bfseries 66} (1991) 683}.

\bibitem{Balberg:2002ue}
S.~Balberg, S.L.~Shapiro and S.~Inagaki, \emph{{Selfinteracting dark matter
  halos and the gravothermal catastrophe}},
  \href{https://doi.org/10.1086/339038}{\emph{Astrophys. J.} {\bfseries 568}
  (2002) 475} [\href{https://arxiv.org/abs/astro-ph/0110561}{{\ttfamily
  astro-ph/0110561}}].

\bibitem{1980MNRAS.191..483L}
D.~{Lynden-Bell} and P.P.~{Eggleton}, \emph{{On the consequences of the
  gravothermal catastrophe}},
  \href{https://doi.org/10.1093/mnras/191.3.483}{\emph{Monthly Notices of the
  Royal Astronomical Society} {\bfseries 191} (1980) 483}.

\bibitem{Koda:2011yb}
J.~Koda and P.R.~Shapiro, \emph{{Gravothermal collapse of isolated
  self-interacting dark matter haloes: N-body simulation versus the fluid
  model}}, \href{https://doi.org/10.1111/j.1365-2966.2011.18684.x}{\emph{Mon.
  Not. Roy. Astron. Soc.} {\bfseries 415} (2011) 1125}
  [\href{https://arxiv.org/abs/1101.3097}{{\ttfamily 1101.3097}}].

\bibitem{17b23f5522334b1ca8bb575ecaf5c01e}
T.~Baumgarte and S.~Shapiro, \emph{Numerical relativity: Solving Einstein's
  equations on the computer}, Cambridge University Press, United Kingdom (Jan.,
  2010),
  \href{https://doi.org/10.1017/CBO9781139193344}{10.1017/CBO9781139193344}.

\bibitem{Bahcall:1976aa}
J.N.~Bahcall and R.A.~Wolf, \emph{{Star distribution around a massive black
  hole in a globular cluster}},
  \href{https://doi.org/10.1086/154711}{\emph{Astrophys. J.} {\bfseries 209}
  (1976) 214}.

\bibitem{1981phki.book.....L}
E.M.~{Lifshitz} and L.P.~{Pitaevskii}, \emph{{Physical kinetics}} (1981).

\bibitem{Sadeghian:2013laa}
L.~Sadeghian, F.~Ferrer and C.M.~Will, \emph{{Dark matter distributions around
  massive black holes: A general relativistic analysis}},
  \href{https://doi.org/10.1103/PhysRevD.88.063522}{\emph{Phys. Rev. D}
  {\bfseries 88} (2013) 063522}
  [\href{https://arxiv.org/abs/1305.2619}{{\ttfamily 1305.2619}}].

\bibitem{1983bhwd.book.....S}
S.L.~{Shapiro} and S.A.~{Teukolsky}, \emph{{Black holes, white dwarfs, and
  neutron stars : the physics of compact objects}} (1983).

\bibitem{Kormendy:2013dxa}
J.~Kormendy and L.C.~Ho, \emph{{Coevolution (Or Not) of Supermassive Black
  Holes and Host Galaxies}},
  \href{https://doi.org/10.1146/annurev-astro-082708-101811}{\emph{Ann. Rev.
  Astron. Astrophys.} {\bfseries 51} (2013) 511}
  [\href{https://arxiv.org/abs/1304.7762}{{\ttfamily 1304.7762}}].

\bibitem{vandenBosch:2016ild}
R.~van~den Bosch, \emph{{Unification of the Fundamental Plane and Super-Massive
  Black Holes Masses}},
  \href{https://doi.org/10.3847/0004-637X/831/2/134}{\emph{Astrophys. J.}
  {\bfseries 831} (2016) 134}
  [\href{https://arxiv.org/abs/1606.01246}{{\ttfamily 1606.01246}}].

\bibitem{Yu:2002sq}
Q.-j.~Yu and S.~Tremaine, \emph{{Observational constraints on growth of massive
  black holes}},
  \href{https://doi.org/10.1046/j.1365-8711.2002.05532.x}{\emph{Mon. Not. Roy.
  Astron. Soc.} {\bfseries 335} (2002) 965}
  [\href{https://arxiv.org/abs/astro-ph/0203082}{{\ttfamily
  astro-ph/0203082}}].

\bibitem{Shankar:2007zg}
F.~Shankar, D.H.~Weinberg and J.~Miralda-Escude, \emph{{Self-Consistent Models
  of the AGN and Black Hole Populations: Duty Cycles, Accretion Rates, and the
  Mean Radiative Efficiency}},
  \href{https://doi.org/10.1088/0004-637X/690/1/20}{\emph{Astrophys. J.}
  {\bfseries 690} (2009) 20} [\href{https://arxiv.org/abs/0710.4488}{{\ttfamily
  0710.4488}}].

\bibitem{Bertschinger:1985pd}
E.~Bertschinger, \emph{{Self - similar secondary infall and accretion in an
  Einstein-de Sitter universe}},
  \href{https://doi.org/10.1086/191028}{\emph{Astrophys. J. Suppl.} {\bfseries
  58} (1985) 39}.

\bibitem{Tambalo:2022wlm}
G.~Tambalo, M.~Zumalac\'arregui, L.~Dai and M.H.-Y.~Cheung,
  \emph{{Gravitational wave lensing as a probe of halo properties and dark
  matter}},  \href{https://arxiv.org/abs/2212.11960}{{\ttfamily 2212.11960}}.

\bibitem{Qiao:2022nic}
C.-K.~Qiao and M.~Zhou, \emph{{Weak Gravitational Lensing of Schwarzschild and
  Charged Black Holes Embedded in Perfect Fluid Dark Matter Halo}},
  \href{https://arxiv.org/abs/2212.13311}{{\ttfamily 2212.13311}}.

\bibitem{Savastano:2022jjv}
S.~Savastano, F.~Vernizzi and M.~Zumalac\'arregui, \emph{{Through the lens of
  Sgr A$^*$: identifying strongly lensed Continuous Gravitational Waves beyond
  the Einstein radius}},  \href{https://arxiv.org/abs/2212.14697}{{\ttfamily
  2212.14697}}.

\bibitem{Oancea:2022szu}
M.A.~Oancea, R.~Stiskalek and M.~Zumalac\'arregui, \emph{{From the gates of the
  abyss: Frequency- and polarization-dependent lensing of gravitational waves
  in strong gravitational fields}},
  \href{https://arxiv.org/abs/2209.06459}{{\ttfamily 2209.06459}}.

\end{thebibliography}\endgroup
\end{document}